\documentclass[journal]{IEEEtran}

\ifCLASSINFOpdf
 
\ifCLASSOPTIONcompsoc
  \usepackage[caption=false,font=normalsize,labelfont=sf,textfont=sf]{subfig}
\else
  \usepackage[caption=false,font=footnotesize]{subfig}
\fi

\usepackage{mathtools}
\usepackage{amsmath}
\usepackage{algorithm,algpseudocode}
\usepackage{hyperref}
\usepackage{amsfonts,bm, dsfont, amssymb, amsthm, verbatim}
\usepackage{enumitem}
\usepackage{graphicx}
\usepackage[font={small,it}]{caption}
\usepackage{tikz}
\usepackage{multicol}

\hypersetup{hidelinks}

\algnewcommand{\Inputs}[1]{%
  \State \textbf{Inputs:}
  \Statex \hspace*{\algorithmicindent}\parbox[t]{.8\linewidth}{\raggedright #1}
}
\algnewcommand{\Initialize}[1]{%
  \State \textbf{Initialize:}
  \Statex \hspace*{\algorithmicindent}\parbox[t]{.8\linewidth}{\raggedright #1}
}

\newcommand{\ma}[1]{\ensuremath{\mathsf{#1}}}
\renewcommand{\vec}[1]{\ensuremath{\mathbf{#1}}}

\newtheorem{proposition}{Proposition}

\newcommand{\ie}{\textit{i.e.}}
\newcommand{\eg}{\textit{e.g.}}
\newcommand{\etal}{\textit{et al.}}
\newcommand\nt[1]{\textcolor{black}{#1}}

\newcommand\rwchng[1]{\textcolor{black}{#1}}
\newcommand\mrw[1]{\textcolor{black}{#1}}

\DeclareMathOperator*{\argmin}{\mathrm{argmin}}
\DeclareMathOperator*{\argmax}{\mathrm{argmax}}
\DeclareMathOperator*{\Var}{Var}
\DeclareMathOperator*{\diag}{diag}
\DeclareMathOperator*{\tr}{tr}
\DeclareMathOperator*{\abs}{abs}

\algdef{SE}[DOWHILE]{Do}{doWhile}{\algorithmicdo}[1]{\algorithmicwhile\ #1}%

\begin{document}

\title{Graph Tikhonov Regularization and Interpolation via Random Spanning Forests}

\author{Yusuf Yi\u{g}it Pilavc{\i}, Pierre-Olivier Amblard, Simon Barthelm\'e, Nicolas Tremblay
\thanks{ This work was partly funded by
	the French National Research Agency in the framework of the "Investissements d’avenir” program (ANR-15-IDEX-02), the LabEx PERSYVAL (ANR-11-LABX-0025-01), the ANR GraVa (ANR-18-CE40-0005)
	the ANR GenGP  (ANR-16-CE23-0008)
	the Grenoble Data Institute (ANR-15-IDEX-02)
	the MIAI@Grenoble Alpes chairs ``LargeDATA at UGA." and ``Pollutants'' (ANR-19-P3IA-0003).}
\thanks{Yusuf Yi\u{g}it Pilavc{\i}, Pierre-Olivier Amblard, Simon Barthelm\'e and Nicolas Tremblay are with CNRS, Univ. Grenoble Alpes, Grenoble INP, GIPSA-lab, Grenoble, France. (e-mail: firstname.lastname@gipsa-lab.fr with firstname.lastname@ =
yusuf-yigit.pilavci@, pierre-olivier.amblard@,  simon.barthelme@, nicolas.tremblay@)}
\thanks{This paper has supplementary downloadable material available at http://ieeexplore.ieee.org., provided by the authors. The material includes proof of proposition \ref{prop:labelprop} and an illustration for Section~\ref{sec:irls}. Contact Yusuf Yi\u{g}it Pilavc{\i} for further questions about this work.}
}


\maketitle

\begin{abstract}
Novel Monte Carlo estimators are proposed to solve both the Tikhonov
regularization (TR) and the interpolation problems on graphs. 
These estimators are based on random spanning forests
(RSF), the theoretical properties of which enable to analyze the estimators' theoretical mean and variance.  
We also show how to perform hyperparameter tuning for these RSF-based
estimators. TR is a component in many well-known algorithms, and we show how the proposed estimators can be easily adapted to avoid expensive intermediate steps in generalized semi-supervised learning, label propagation, Newton's method and iteratively reweighted least squares. 
In the experiments, we illustrate the proposed methods on several problems and provide observations on their run time.
\end{abstract}


\IEEEpeerreviewmaketitle

\section{Introduction}

\IEEEPARstart{G}{raphs} are ubiquitous models of complex structures, \eg~social,
transportation, sensors or neuronal networks. 
The vertices and edges, the main components of graphs, are natural representations of the elements of a network
and the links between them, respectively. In many applications, these networks often come with
data on the elements. 
For example, in a transportation network 
\nt{(the roads and their intersections  are respectively the nodes and edges),
the data can be the traffic measured at the intersections}; or in a brain network, it could be the activity of each individual brain region~\cite{Huang2018}. 
Such type of data over vertices are called graph signals~\cite{Shuman2013, Sandryhaila2014}.  
\\

\begin{figure}
	\includegraphics[width=9cm,height=5cm]{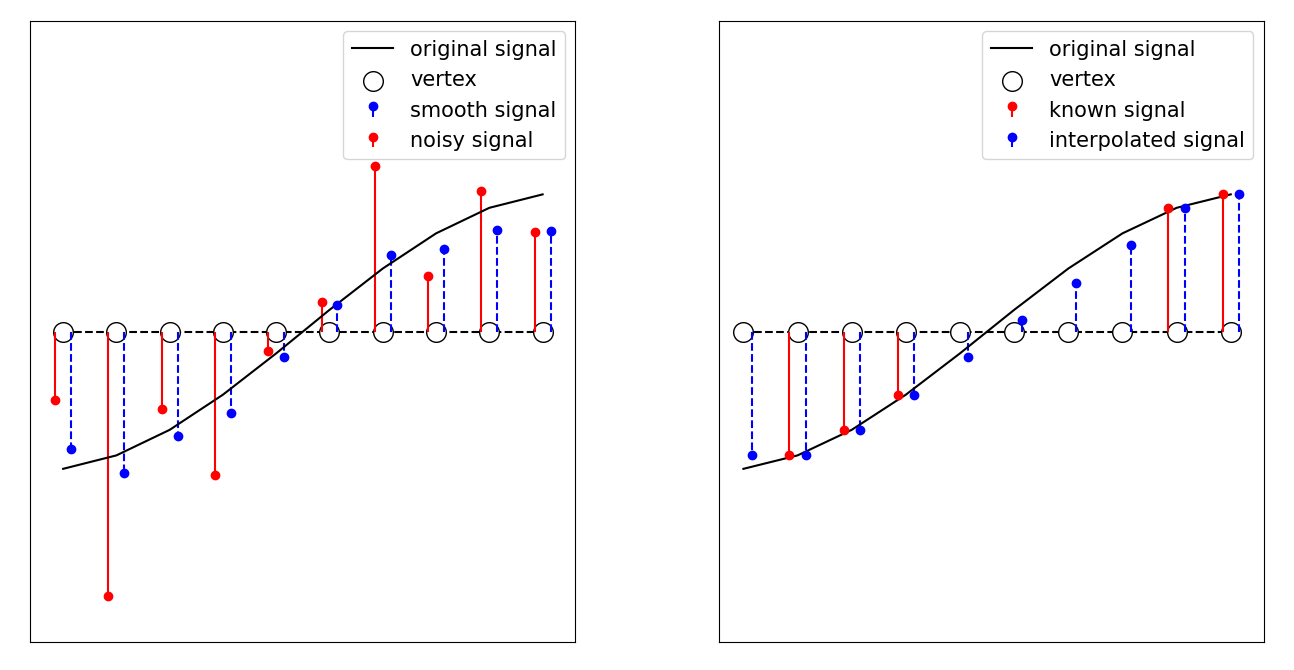}
	\caption{
		\rwchng{Smoothing (left) and interpolation (right) in discrete \nt{1D} signal processing can be considered as a graph signal processing problem where the underlying graph is a path graph.}}
	\label{fig:introfig}
\end{figure}

\noindent\rwchng{\textbf{Graph Tikhonov regularization and interpolation. } Graph signal processing (GSP) is a field of research developed in the last decade dedicated to extending classical signal processing tools to graph signals~\cite{Shuman2013,Ortega2018}. 
In this paper, we focus on two essential operations on graph signals, namely graph Tikhonov regularization and graph interpolation. 
In the former, noisy measurements of a graph signal are known over all vertices. Tikhonov regularization is a smoothing operation and recovers the underlying signal assuming that it varies slowly between neighboring vertices in the graph. 
Under the same assumption, interpolation, on the other hand, is the operation of recovering the signal when it is known only over a few vertices (see Fig.~\ref{fig:introfig} for an illustration on the path graph). 
Both operations are thoroughly analyzed in ~\cite{zhu_2002learning,Smola2003,Belkin2004,Pesenson2011} and we refer the reader to Section~\ref{sec:estimators} for formal definitions. 
}

\noindent \rwchng{\textbf{Hyperparameter tuning.} As usual in regularization schemes, both graph TR and interpolation have hyper-parameters that need to be to tuned. Automatically tuning these parameters is not a trivial problem. In the classical framework of linear smoothers~\cite{hastie_elements_2005}, there exists many methods such as Akaike and Bayesian information criteria, Marlow's Cp, leave-one-out and generalized cross validations or Stein's unbiased risk estimator. In this paper, we study possible graph adaptations for some of them to automatically tune hyper-parameters in graph TR and interpolation.   } \\

\noindent\textbf{\rwchng{Solving graph TR and interpolation as a building block}}.  \rwchng{The solutions of graph TR and interpolation appear in the intermediate steps of various graph algorithms.}

\rwchng{Semi-supervised node classification is one problem that naturally boils down to a graph interpolation scheme. 
A few labels are a priori known and the purpose is to infer, via graph interpolation, the labels for all nodes given the structure of the underlying graph.
One solution to this problem is given by a well-known Semi-Supervised Learning (SSL) algorithm, called label propagation\mrw{~\cite{zhu_2002learning,Zhu2005}}. Label propagation is an iterative algorithm whose solution converges to the solution of the graph interpolation problem. 
Along with the connections to the Dirichlet boundary problem on
graphs~\cite{Grady2003}, this solution can be seen as assigning labels to nodes that \emph{smoothly} interpolate between known labels at the boundary points. 
}

Graph TR, besides its use for graph signal denoising, also appears in SSL algorithms for node classification, such as in the work of Zhou~\etal~\cite{zhou2004learning}, later generalized by Avrachenkov~\etal~\cite{avrachenkov2012generalized}. Moreover, graph TR \mrw{can also be} used in graph optimization algorithms. 
Two examples are Newton's method~\cite{boyd_convex_2004} and iteratively reweighted least squares (IRLS)~\cite{Burrus1994}. 
In both methods, the computationally expensive steps can be formulated and solved as graph TR problems.
\\

\noindent \rwchng{\textbf{Classical approaches.} Both graph TR and interpolation have closed-form solutions. 
However, directly computing these solutions requires the inversion of a matrix
of size $n\times n$, and thus, up to $\mathcal{O}(n^3)$ elementary operations.
For large graphs (\ie~with $n \geq 10^4$), this is prohibitive and the
state-of-the-art relies on approximate methods. These approaches may be roughly
separated in two groups, iterative methods (\eg~conjugate gradient method with
preconditioning \cite{Saad}) and polynomial approximations (\eg~Chebyshev
polynomials \cite{Shuman2011}). \rwchng{Both classes} of methods run in linear
time with the number of edges $m$. } \\ 

\noindent \textbf{Random processes on graphs.} A longstanding and fruitful
approach to studying the properties of graphs has been to study the properties of random processes
on graphs (via random walks, for instance). This paper will take such a perspective to propose novel estimators for the two problems presented.

For instance, a very well-known fact is the link between the smallest non-null
eigenvalue of the graph's Laplacian matrix and the mixing time of a random walk on the
graph (see, \textit{e.g.,}~\cite{chung1997spectral}). Other examples include
properties of electrical networks, such as potential functions or effective resistances, that can be interpreted in terms of probabilistic quantities defined for random walks such as hitting time probabilities~\cite{grimmett2018probability}. 

\rwchng{Closer to our work, Wu~\etal~\cite{Wu2012,Wu2016} study probabilistic properties of a particular random process, called partially absorbed random walk; and further leverage random walks to give practical insights into algorithms dedicated to image retrieval, local clustering and semi-supervised learning.\\
}

\noindent\textbf{Random spanning forests.} 
In this paper, we will focus on random spanning forests (RSFs): random
collections of disjoint trees that cover all nodes in the graph (a formal
definition  is in \rwchng{Section}~\ref{sec:backgroundonRSF}).
\rwchng{Interestingly, connections between graph TR / interpolation and RSFs have been observed by a few authors in the past, such as in~\cite{Avrachenkov2017}}. 
Avena~\etal~\cite{Avena2013,avena_two_2017} analyze several aspects of \rwchng{these connections}. RSFs not only have a rich theoretical background (they have connections with Markov chains, determinantal point processes, spectral graph theory), they also come with an efficient sampling algorithm~\cite{Avena2013}: a variant of Wilson's algorithm~\cite{Wilson1996} based on loop-erased random walks. 
\\

\noindent \textbf{Our contributions}. \rwchng{In the conference paper~\cite{pilavci2020}, we have already presented some preliminary results. Differing from~\cite{pilavci2020}, in this work,}
    \rwchng{
    \begin{itemize}
    \item We extend the theoretical analysis on the RSF-based estimators proposed in~\cite{pilavci2020}:   
    	\begin{itemize}
    	\item Building upon known results on RSFs, we generalize the error analysis to cases where the regularization hyperparameter is not constant over the vertices.
    	\item We provide a precise expression for the computational cost of the estimators. 
	    \end{itemize}	 
   	\item Using certain statistics of RSFs, we provide a novel scheme to correctly tune the hyperparameters.
   	\item We discuss how these estimators may be leveraged to approximately solve the graph interpolation problem. 
	\item \mrw{We show that the proposed estimators have versatile use in other graph-based problems and their solutions such as generalized semi-supervised learning, label propagation, Newton's method and IRLS.
	\item \mrw{We provide illustrations for different use cases. Tikhonov regularization and Newton's method are demonstrated in image denoising applications. Graph interpolation and generalized SSL algorithms are used to solve classification tasks on citation networks.
	\item \mrw{The runtime of the proposed methods are compared with state-of-the-art methods on certain graphs. } }
	}
   	\end{itemize}
	}
\rwchng{The Julia code to reproduce this paper's results is available on the authors' website.\footnote{\url{https://gricad-gitlab.univ-grenoble-alpes.fr/pilavciy/rsf_sipn_codes.git}}}
\\

\noindent \textbf{Organization of the paper.} We start with the necessary background on graphs and RSFs in Section \ref{sec:backgroundonRSF}. Then, we introduce the proposed methods in Section~\ref{sec:estimators}. In Section~\ref{sec:graphotherprobs}, we examine several extensions to different graph-related problems. Finally, in Section~\ref{sec:experiments}, we illustrate the methods \mrw{and provide a runtime analysis} and we conclude in Section~\ref{sec:conclusion}.

\section{Background on RSFs}
\label{sec:backgroundonRSF}
This section contains background on graph theory, random spanning trees and forests. 
\subsection{Graph theory} 
A directed weighted graph $\mathcal{G} = (\mathcal{V},\mathcal{E},w)$ consists of a set of vertices $
\mathcal{V}= \{1,2,...,n\}$ and edges $\mathcal{E}= \{(i,j) \in
\mathcal{V}\times\mathcal{V}\}$. The weight function $w:
\mathcal{V}\times\mathcal{V} \rightarrow \mathbb{R}^+$ maps each edge in
$\mathcal{E}$ to a positive weight and others to 0. A graph is called undirected
if $w(i,j) = w(j,i)$ for all distinct vertices $i$ and $j$. In the following,
unless otherwise specified only undirected graphs are considered. 
Graphs are often represented using matrices, and several matrix descriptions are available.

The weighted adjacency matrix is $\ma{W} = [w(i,j)]_{i,j} \in \mathbb{R}^{n\times n}$. The degree matrix is the diagonal matrix $\ma{D}\in \mathbb{R}^{n\times n}$ with $\ma{D}_{i,i} = \sum_{j\in\mathcal{N}(i)}{w(i,j)}$ and $\mathcal{N}(i)$ is the set of nodes connected to $i$. The graph Laplacian matrix is defined as $\ma{L} = \ma{D} - \ma{W}$. It is semi-positive definite~\cite{chung1997spectral} and its eigenvalues and eigenvectors are denoted by $0=\lambda_1\leq\lambda_2\leq\hdots\leq\lambda_n $ and $\ma{U} = (\vec{u}_1|\vec{u}_2|\hdots|\vec{u}_n)$, respectively. 
The multiplicity of eigenvalue $0$ equals the number of connected components in the graph~\cite{chung1997spectral}. For undirected graphs, all of these square matrices are symmetric. 

Another way to represent graphs is via the edge incidence matrix $\ma{B} = (\vec{b}_1 | \vec{b}_2 | \hdots |\vec{b}_m)^\top \in \mathbb{R}^{m \times n}$ where $\vec{b}_k\in\mathbb{R}^n$ is a vector associated to the $k$-th edge $(i,j)$. The only nonzero entries of $ \vec{b}_k$ are $\vec{b}_{k}(i) = \pm\sqrt{w(i,j)} $ and $\vec{b}_{k}(j) = \mp\sqrt{w(i,j)} $. In undirected graphs, the signs of the non-zero entries of $\vec{b}_k$ can be arbitrarily chosen as long as they are opposite. Although this matrix seems less natural than the others, it often appears in graph theory. One example is the well known identity $\ma{L} = \ma{B}^\top\ma{B}$.
\begin{figure*}[tb]
	\centering
	\begin{tikzpicture}[ 
	every edge/.style = {draw=black,very thick},
	vrtx/.style args = {#1/#2}{%
		circle, draw, thick, fill=white,
		minimum size=2mm,scale=1, label=#1:#2}
	]
	\node(A) [vrtx=above/] at (0.5, 1) {};
	\node(B) [vrtx=left/] at (0, 0) {};
	\node(C) [vrtx=right/] at (1,0) {};
	\node(D) [vrtx=right/] at (1.5,-1) {};
	\node(E) [vrtx=left/] at (-1.5, -1) {};
	\node(F) [vrtx=below/] at (0, -1) {};
	\node(G) [vrtx=right/] at (1,-2) {};
	\node(H) [vrtx=left/] at (-1,-2) {};
	\draw[-]   
	(A) edge (B)
	(B) edge (C)
	(A) edge (C)
	(C) edge (D)
	(F) edge (D)
	(B) edge (F)
	(G) edge (D)
	(F) edge (H)
	(E) edge (H)
	(E) edge (F)
	(G) edge (F)
	(B) edge (E);
	\end{tikzpicture}
	\begin{tikzpicture}[
	every edge/.style = {draw=black,very thick},
	vrtx/.style args = {#1/#2}{%
		circle, draw, thick, fill=white,
		minimum size=2mm, label=#1:#2,scale = 1}
	]

	\node(A) [vrtx=above/] at (0.5, 1) {};
	\node(B) [vrtx=left/] at (0, 0) {};
	\node(C) [vrtx=right/] at (1,0) {};
	\node(D) [vrtx=right/] at (1.5,-1) {};
	\node(E) [vrtx=left/] at (-1.5, -1) {};
	\node(F) [vrtx=below/] at (0, -1) {};
	\node(G) [vrtx=right/] at (1,-2) {};
	\node(H) [vrtx=left/] at (-1,-2) {};
	
	\draw[-]   (A) edge (B)
	(A) edge (C)
	(B) edge (E)
	(C) edge (D)
	(E) edge (H)
	(G) edge (F)
	(B) edge (F);
	\end{tikzpicture}
	\begin{tikzpicture}[ 
	every edge/.style = {draw=black,very thick},
	vrtx/.style args = {#1/#2}{%
		circle, draw, thick, fill=white,
		minimum size=2mm,scale=1, label=#1:#2}
	]
	\node(A) [vrtx=above/,fill=red] at (0.5, 1) {};
	\node(B) [vrtx=left/] at (0, 0) {};
	\node(C) [vrtx=right/] at (1,0) {};
	\node(D) [vrtx=right/] at (1.5,-1) {};
	\node(E) [vrtx=left/] at (-1.5, -1) {};
	\node(F) [vrtx=below/] at (0, -1) {};
	\node(G) [vrtx=right/] at (1,-2) {};
	\node(H) [vrtx=left/] at (-1,-2) {};

	\draw[->]   
	(B) edge (A)
	(C) edge (A)
	(E) edge (B)
	(H) edge (E)
	(F) edge (B)
	(G) edge (F)
	(D) edge (C);
	\end{tikzpicture}
	\begin{tikzpicture}[
	every edge/.style = {draw=black,very thick},
	vrtx/.style args = {#1/#2}{%
		circle, draw, thick, fill=white,
		minimum size=2mm, label=#1:#2,scale = 1}
	]
	\node(A) [vrtx=above/,fill=red] at (0.5, 1) {};
	\node(B) [vrtx=left/] at (0, 0) {};
	\node(C) [vrtx=right/] at (1,0) {};
	\node(D) [vrtx=right/] at (1.5,-1) {};
	\node(E) [vrtx=left/,fill=red] at (-1.5, -1) {};
	\node(F) [vrtx=below/,fill=red] at (0, -1) {};
	\node(G) [vrtx=right/] at (1,-2) {};
	\node(H) [vrtx=left/] at (-1,-2) {};
	
	\draw[->]   
	(B) edge (A)
	(C) edge (A)
	(H) edge (E)
	(D) edge (C)
	(G) edge (F);
	\end{tikzpicture}    
	\caption{From left to right, a graph $\mathcal{G}$, a spanning tree on $\mathcal{G}$, a rooted spanning tree on $\mathcal{G}$ and a rooted spanning forest on $\mathcal{G}$ (roots are colored in red)}
	\label{fig:graphexamples}
\end{figure*}
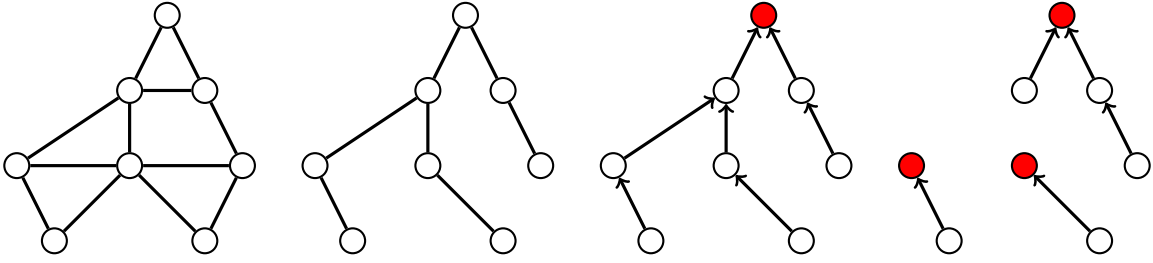
\subsection{Random spanning trees and Wilson's algorithm}

Let us recall the definition of random spanning trees (RSTs). Consider a graph $\mathcal{G} = (\mathcal{V},\mathcal{E},w)$. 
A subgraph of $\mathcal{G}$ is a graph whose vertex and edge sets are subsets of $\mathcal{V}$ and $\mathcal{E}$, respectively, and its edge weights are valued by $w$. A subgraph contains a cycle whenever there exists a pair of vertices $(u,v)$ that are connected via (strictly) more than one path. 
If there exists no such pair, the subgraph is called a tree.
A spanning tree $\tau = (\mathcal{V}_\tau,\mathcal{E}_\tau,w_\tau)$ is a tree
whose vertex set $\mathcal{V}_\tau$ is equal to $\mathcal{V}$.  A rooted spanning tree $\tau_r$ is
a directed spanning tree where all edges are directed towards a node called the root. See Fig. \ref{fig:graphexamples} for illustrations. 

Our work is related to a particular distribution on spanning trees called random spanning trees (RSTs). An RST $T$ is a randomly generated spanning tree from the following distribution:  
\begin{equation}
 \mathbb{P}(T = \tau) \propto \prod_{(i,j) \in \tau}w(i,j).
 \label{eq:weightedspanningtree}
\end{equation}
Note that this distribution becomes uniform over all possible spanning trees whenever the given graph is unweighted~\ie~$\forall i,j\in\mathcal{V}\text{, }w(i,j)\in\{0,1\}$: 
 \begin{equation}
 \mathbb{P}(T = \tau) = \frac{1}{|\mathcal{T}|}
 \label{eq:UST}
 \end{equation} 
where $\mathcal{T}$ is the set of all spanning trees. In this particular case, the random tree $T$ is also known as a uniform spanning tree (UST) in the literature. 

In his celebrated work~\cite{Wilson1996}, Wilson proposes an algorithm, called \texttt{RandomTreeWithRoot}, that samples a random spanning tree from the set of all spanning trees rooted in node $r$. Wilson also shows that, in the case of undirected graphs, sampling an unrooted RST amounts to: i/~choosing uniformly a root, ii/~running \texttt{RandomTreeWithRoot}, and iii/~erasing the orientation.

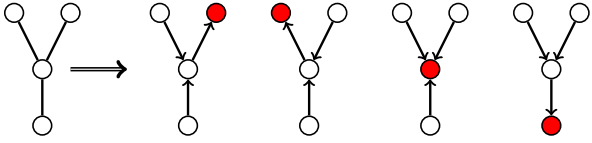
\begin{figure}
\centering
	\scalebox{0.75}{
		\begin{tikzpicture}[ 
            every edge/.style = {draw=black,very thick},
             vrtx/.style args = {#1/#2}{%
              circle, draw, thick, fill=white,
              minimum size=2mm,scale=1, label=#1:#2},
            baseline=(current bounding box.center)]
        \node(A) [vrtx=above/] at (1, 1) {};
        \node(B) [vrtx=left/] at (0, 1) {};
        \node(C) [vrtx=right/] at (0.5,0) {};
        \node(D) [vrtx=right/] at (0.5,-1) {};
        	
		\draw[-]   
        (A) edge (C)
        (B) edge (C)
        (D) edge (C);
        \draw[->,double,thick] (1.0,0) -- (2,0); 
        \end{tikzpicture}
        \begin{tikzpicture}[ 
            every edge/.style = {draw=black,very thick},
             vrtx/.style args = {#1/#2}{%
              circle, draw, thick, fill=white,
              minimum size=2mm,scale=1, label=#1:#2},baseline=(current bounding box.center)
            ]
        \node(A) [vrtx=above/,fill=red] at (1, 1) {};
        \node(B) [vrtx=left/] at (0, 1) {};
        \node(C) [vrtx=right/] at (0.5,0) {};
        \node(D) [vrtx=right/] at (0.5,-1) {};
        	
		\draw[->]   
        (C) edge (A)
        (B) edge (C)
        (D) edge (C);
        \end{tikzpicture}
        \hspace{0.3cm}
		\begin{tikzpicture}[ 
            every edge/.style = {draw=black,very thick},
             vrtx/.style args = {#1/#2}{%
              circle, draw, thick, fill=white,
              minimum size=2mm,scale=1, label=#1:#2},baseline=(current bounding box.center)
            ]

        \node(A) [vrtx=above/] at (1, 1) {};
        \node(B) [vrtx=left/,fill=red] at (0, 1) {};
        \node(C) [vrtx=right/] at (0.5,0) {};
        \node(D) [vrtx=right/] at (0.5,-1) {};
        	
		\draw[->]   
        (A) edge (C)
        (C) edge (B)
        (D) edge (C);
        \end{tikzpicture}
        \hspace{0.3cm}
		\begin{tikzpicture}[ 
            every edge/.style = {draw=black,very thick},
             vrtx/.style args = {#1/#2}{%
              circle, draw, thick, fill=white,
              minimum size=2mm,scale=1, label=#1:#2},baseline=(current bounding box.center)
            ]

        \node(A) [vrtx=above/] at (1, 1) {};
        \node(B) [vrtx=left/] at (0, 1) {};
        \node(C) [vrtx=right/,fill=red] at (0.5,0) {};
        \node(D) [vrtx=right/] at (0.5,-1) {};
        	
		\draw[->]   
        (A) edge (C)
        (B) edge (C)
        (D) edge (C);
        \end{tikzpicture}
        \hspace{0.3cm}
		\begin{tikzpicture}[ 
            every edge/.style = {draw=black,very thick},
             vrtx/.style args = {#1/#2}{%
              circle, draw, thick, fill=white,
              minimum size=2mm,scale=1, label=#1:#2},baseline=(current bounding box.center)
            ]

        \node(A) [vrtx=above/] at (1, 1) {};
        \node(B) [vrtx=left/] at (0, 1) {};
        \node(C) [vrtx=right/] at (0.5,0) {};
        \node(D) [vrtx=right/,fill=red] at (0.5,-1) {};
        	
		\draw[->]   
        (A) edge (C)
        (B) edge (C)
        (C) edge (D);
        \end{tikzpicture}
        }
\caption{All possible rooted spanning trees associated with a given undirected spanning tree. For four vertices, four different rooted trees exist.}
\label{fig:uniformroots}
\end{figure}

\subsection{Random spanning forests}
A forest is a set of disjoint trees. When all the trees in a forest are rooted, it is called a rooted forest. A rooted spanning forest, generically denoted by $\phi$, reaches all the vertices in the graph. Let $\rho$ be the function that maps any rooted spanning forests to its set of roots. The number of roots $|\rho(\phi)|$ is between 1 and $n$. For $|\rho(\phi)| = 1$, $\phi$ corresponds to a rooted spanning tree. See Fig. \ref{fig:graphexamples} for illustrations.
~\\
     
\noindent \textbf{Random Spanning Forests}. Let $\mathcal{F}$ be the set of all rooted spanning forests. A random spanning forest (RSF) is a random variable whose outcome space is $\mathcal{F}$. Among many possible options, we focus on the following parametric distribution for RSFs. 
For a fixed parameter $q>0$, $\Phi_q$ is a random variable in $\mathcal{F}$ verifying: 
\begin{equation}
\forall\phi \in \mathcal{F}, \quad \mathbb{P}(\Phi_q = \phi) \propto q^{|\rho(\phi)|}\prod_{(i,j) \in \phi}w(i,j).
\label{eq:rsfdist}
\end{equation} 
An algorithm \cite{avena_two_2017} to sample from this distribution is derived from \texttt{RandomTreeWithRoot}. This algorithm: 
\begin{enumerate}
\item  extends the graph $\mathcal{G} = (\mathcal{V},\mathcal{E},w)$ by adding a node called $\Gamma$.
\item connects each node $i$ in $\mathcal{V}$ to $\Gamma$ with an edge of weight $w(i,\Gamma) = q$.
\item runs \texttt{RandomTreeWithRoot} by setting $\Gamma$ as the root to obtain a spanning tree rooted in $\Gamma$ in the extended graph.
\item deletes the edges incident to $\Gamma$ in the obtained tree to yield a forest in the original graph. 
\end{enumerate} 
The result is a rooted spanning forest whose root set is formed by the nodes which were neighbors of $\Gamma$. For every distinct spanning tree rooted at $\Gamma$, a distinct spanning forest is obtained after removing the root and its incident edges. Using this one-to-one relation ensures that this algorithm indeed samples from the distribution in \mrw{Eq.}~\eqref{eq:rsfdist}:
\begin{equation}
\begin{split}
	\mathbb{P}(T_\Gamma = \tau_\Gamma)=\mathbb{P}(\Phi_q = \phi) &\propto \prod_{(i,j) \in {\tau_\Gamma}}w(i,j)  \\ &\propto  \prod_{(i,\Gamma)\in \tau_\Gamma}q \prod_{\substack{(i,j) \in {\tau_\Gamma} \\ i,j\not=\Gamma}}w(i,j) \\ &\propto  q^{|\rho(\phi)|}\prod_{(i,j) \in {\phi}}w(i,j).
\end{split}
\end{equation}

An implementation of this algorithm is detailed in
Algorithm~\ref{alg:randomforestalg}. In the algorithm, \texttt{rand} (line 7)
returns a uniform random value between 0 and 1 and \texttt{RandomSuccessor} (line 11) returns a random node $i$ from $\mathcal{N}(u)$ with probability $\frac{w(u,i)}{\sum_{j\in \mathcal{N}(u)}w(u,j)}$. At termination, the array Next contains all the necessary information to build the sampled spanning forest. 

The expected run time of 
\texttt{RandomForest} is the expected number of calls of \texttt{RandomSuccessor} before termination. For \texttt{RandomTreeWithRoot}, the number of calls equals to the mean commute time, \ie, the expected length of a random walk going from node $i$ to $j$ and back (see theorem 2 in~\cite{Wilson1996}). In proposition 1 of~\cite{Marchal2000}, Marchal rewrites this commute time in terms of graph matrices. Adapting his result to the current setting, the expected run time of \texttt{RandomForest} can be shown to equal the trace of $\left((\ma{L}+q\ma{I})^{-1}(\ma{D}+q\ma{I})\right)$, a rough upper-bound of which is $n+2|\mathcal{E}|/q$, which is linear with the number of edges. 
  \\
\begin{algorithm}
    \caption{\texttt{RandomForest}}
    \label{alg:randomforestalg}
    \begin{algorithmic}[1]
        \Inputs{$\mathcal{G}=(\mathcal{V},\mathcal{E},w)$\\
        $q\nt{>0}$}
        \Initialize{
        \textit{\# Initially, the forest is empty}\\	
        $\forall i\in\mathcal{V},\quad\text{InForest}[i] \leftarrow \texttt{false} $ 
        \\ $\forall i\in\mathcal{V}, \quad\text{Next[i]} \leftarrow -1 $ 
         \\ $\forall i\in \mathcal{V}, \quad\text{d}[i] \leftarrow \sum_{j\in\mathcal{N}(i)} w(i,j)$\textit{ \# Degrees} 
   	     } 
        \For{$i \leftarrow \text{1 to }|\mathcal{V}|$}            
            \State{$u \leftarrow i$}
            \State{\textit{\# Start a random walk to create a forest branch}}
			\While{\textbf{not} $\text{InForest}[u]$} \textit{ \# Stop if $u$ is in the forest}
            \If{$\texttt{rand} \leq \frac{q}{q+\text{d}[u] }   $} \textit{\#If true, $u$ becomes a root}
            	\State{$\text{InForest}[u] \leftarrow \texttt{true}$ \textit{ \# Add $u$ to the forest}} 
            	\State{$\text{Next}[u] \leftarrow -1$ \textit{ \# Set next of $u$ to null}} 
            \Else \textit{ \# If false, continue the random walk}
            
                \State{$\text{Next[u]} \leftarrow \texttt{RandomSuccessor}(u,\mathcal{G})$}
                \State{$u\leftarrow\text{Next}[u]$}
			\EndIf
		\EndWhile 

			\State{$u\leftarrow i$ \textit{ \# Go back to the initial node}}
			\State{\textit{\# Add the newly created branch to the forest}}
	        \While{\textbf{not} $\text{InForest}[u]$}
    	        \State{$\text{InForest}[u] \leftarrow \texttt{true}$}
        	    \State{$u\leftarrow\text{Next}[u]$}
        	\EndWhile
        \EndFor
        \State{\texttt{return} Next}
    \end{algorithmic}
\end{algorithm}

\noindent \textbf{Varying $q$ over nodes}. The original graph can also be extended by setting\footnote{\nt{A requirement when choosing the $q_i$'s is that the extended graph stays connected: when $q$ is chosen uniform, it thus has to be chosen strictly positive. When $q$ is not chosen uniform, each $q_i$ can be chosen positive or null as long as the extended graph is connected (for instance, all $q_i$'s cannot be null)}} $w(i,\Gamma) \leftarrow q_i \nt{\geq}0 , \forall i \in \mathcal{V}$, that is, by connecting the added node $\Gamma$ with links of unequal weights. In this case, the distribution of sampled forests becomes:  
\begin{equation}
\mathbb{P}(\Phi_Q = \phi) \propto \prod_{i \in \rho(\phi)}q_i\prod_{(i,j) \in \phi}w(i,j), \quad \phi \in \mathcal{F}
\label{eq:nonhomqrsfdist}
\end{equation}
where $Q = \{q_1,q_2,\hdots,q_n\}$ is the collection of  parameters. Algorithm~\ref{alg:randomforestalg} can  easily be  adapted by modifying the scalar input $q$ to $Q = \{q_1,q_2,\hdots,q_n\}$ and $\frac{q}{q+d[u]}$ to $\frac{q_u}{q_u+d[u]}$ at line 7. In addition, the average run time in this case becomes $\tr\left((\ma{L}+\ma{Q})^{-1}(\ma{D}+\ma{Q})\right)$, with $\ma{Q} = \diag (q_1,\hdots,q_n)$. 
~\\

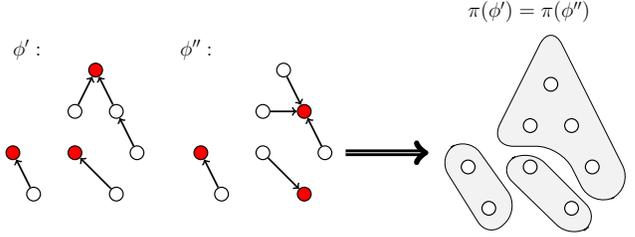
\begin{figure}
	\centering
	\scalebox{0.55}{
 	        \begin{tikzpicture}[
            every edge/.style = {draw=black,very thick},
             vrtx/.style args = {#1/#2}{%
                  circle, draw, thick, fill=white,
                  minimum size=2mm, label=#1:#2}
                ,baseline=(current bounding box.center)]
            
            \node(A) [vrtx=center/,fill=red] at (0.5, 1) {};
            \node(B) [vrtx=center/] at (0, 0) {};
            \node(C) [vrtx=center/] at (1,0) {};
            \node(D) [vrtx=center/] at (1.5,-1) {};
            \node(G) [vrtx=center/] at (1,-2) {};
            \node(F) [vrtx=center/,fill=red] at (0, -1) {};
            \node(H) [vrtx=center/] at (-1,-2) {};
            \node(E) [vrtx=center/,fill=red] at (-1.5, -1) {};
            \node[text width=2cm,font=\Large]  at (-0.5,1.5) {$\phi':$};
            
            \draw[->] (B) edge (A)
            (C) edge (A) 
            (D) edge (C)
            (G) edge (F)
            (H) edge (E);
            \end{tikzpicture}
            \hspace{0.5cm}
            \begin{tikzpicture}[
            every edge/.style = {draw=black,very thick},
             vrtx/.style args = {#1/#2}{%
                  circle, draw, thick, fill=white,
                  minimum size=2mm, label=#1:#2}
                ,baseline=(current bounding box.center)]
            
            \node(A) [vrtx=center/] at (0.5, 1) {};
            \node(B) [vrtx=center/] at (0, 0) {};
            \node(C) [vrtx=center/,fill=red,fill=red] at (1,0) {};
            \node(D) [vrtx=center/] at (1.5,-1) {};
            \node(G) [vrtx=center/,fill=red] at (1,-2) {};
            \node(F) [vrtx=center/] at (0, -1) {};
            \node(H) [vrtx=center/] at (-1,-2) {};
            \node(E) [vrtx=center/,fill=red] at (-1.5, -1) {};
            \node[text width=4cm,font=\Large]  at (0,1.5)  {$\phi'':$};
            \draw[->] (B) edge (C)
            (A) edge (C) 
            (D) edge (C)
            (F) edge (G)
            (H) edge (E);
            \draw[->,double,line width = 2] (2,-1) -- (4,-1);
          
            \end{tikzpicture}
            \begin{tikzpicture}[
            every edge/.style = {draw=black,very thick},
             vrtx/.style args = {#1/#2}{%
                  circle, draw, thick, fill=white,
                  minimum size=2mm, label=#1:#2}
                ,baseline=(current bounding box.center)]
            
             \draw [rounded corners=6mm,fill=gray!10] (0.5,2.5) -- (-1,-0.5)-- (0.75,-0.5)  --(1.5,-2)--  (2.5,-1.5)--    cycle;
            \node(A) [vrtx=center/] at (0.5, 1) {};
            \node(B) [vrtx=center/] at (0, 0) {};
            \node(C) [vrtx=center/] at (1,0) {};
            \node(D) [vrtx=center/] at (1.5,-1) {};
             \draw [rounded corners=5mm,fill=gray!10] (0,-0.55) -- (1.7,-2.25) -- (1,-2.75) -- (-0.8,-1) -- cycle;        
            \node(G) [vrtx=center/] at (1,-2) {};
            \node(F) [vrtx=center/] at (0, -1) {};
            \draw [rounded corners=5mm,fill=gray!10] (-1.5,-0.25) -- (-2.25,-0.75) -- (-1.00,-2.75)-- (-0.25,-2.) -- cycle;
            \node(H) [vrtx=center/] at (-1,-2) {};
            \node(E) [vrtx=center/] at (-1.5, -1) {};
                         \node[text width=4cm,font=\Large]  at (0.5,2.75) {$\pi(\phi')=\pi(\phi'')$};

            \end{tikzpicture}
            }
            \caption{Two different rooted spanning forests (on the left) with the same corresponding partition (on the right) }
            \label{fig:partition}
\end{figure}

\noindent \textbf{Random partitions}. A partition of $\mathcal{V}$, denoted by $\mathcal{P}$, is a set of disjoint subsets whose union equals $\mathcal{V}$. The trees of $\Phi_q$ give a random partition of $\mathcal{V}$ by splitting it into $|\rho(\Phi_q)|$ disjoint subsets. Let us enumerate the trees from $1$ to $|\rho(\Phi_q)|$ and denote the vertex set of the $k$-th tree as $\mathcal{V}_k\subset\mathcal{V}$. Let $\pi$ be a function that outputs the partition for a given spanning forest. Then, the random partition of $\mathcal{V}$ derived from $\Phi_q$ is $\pi(\Phi_q)=(\mathcal{V}_1,\hdots,\mathcal{V}_{|\rho(\Phi_q)|})$ with $|\pi(\Phi_q)|$ subsets. Note that this function is a many-to-one mapping because different spanning forests may correspond to the same partition (see Figure~\ref{fig:partition}). 
 
\subsection{Useful properties of $\Phi_q$}
Recent studies in~\cite{Avena2013,avena_two_2017} have established some theoretical properties of $\Phi_q$ that we reproduce here. ~\\

 \noindent \textbf{The root process.} To start with, Proposition 2.2 in\mrw{~\cite{avena_two_2017}} states that $\rho(\Phi_q)$ is sampled from a determinantal point process (DPP)~\cite{Kulesza2012} with marginal kernel:
\begin{equation}
 \ma{K}= (q\ma{I} + \ma{L})^{-1}q\ma{I}.
 \label{eq:rootkernel}
\end{equation}
This means that the inclusion probabilities verify: 
\[ \forall \mathcal{S}  \subset \mathcal{V}, \quad \mathbb{P}(\mathcal{S} \in \rho(\Phi_q)) = \det \ma{K}_{\mathcal{S} } \]
where $\ma{K}_{\mathcal{S} } = [\ma{K}_{i,j}|(i,j)\in\mathcal{S} \times\mathcal{S} ]$ is the submatrix of $\ma{K}$ reduced to the rows and columns indexed by $\mathcal{S}$. 
~\\

\noindent \textbf{Cardinality of $\rho(\Phi_q)$}. As a consequence of $\rho(\Phi_q)$ being a DPP, the first two moments of $|\rho(\Phi_q)|$ verify~\cite{Kulesza2012}: 
\begin{equation}
\begin{split}
	&\mathbb{E}[|\rho(\Phi_q)|] = \tr(\ma{K}) = \sum_{i}\frac{q}{q+\lambda_i}, \\ &\Var(|\rho(\Phi_q)|) = \tr(\ma{K} -\ma{K}^2)= \sum_{i} \frac{\lambda_i q}{(q+\lambda_i)^2}
	\label{eq:cardinalityrho}	
\end{split}
\end{equation}
where the $\lambda_i$'s are the eigenvalues of $\ma{L}$.
~\\

\noindent \textbf{The root probability distribution}. Given any rooted spanning forest $\phi$, define the \emph{root function} $r_{\phi}: \mathcal{V} \rightarrow \rho(\phi)$ which takes as input any node $i$ and outputs the root of the tree which $i$ belongs to. In \cite{Avena2013,avena_two_2017}, the authors show that the probability, for any node pair $(i,j)$, that $i$ is rooted in $j$ reads:
\begin{equation}
	 \forall i,j \in \mathcal{V} \quad \mathbb{P}(r_{\Phi_q}(i) = j) = \ma{K}_{ij}. 
	\label{eq:probij}
\end{equation}       
~\\

\noindent \textbf{Conditioning on a partition}. Let $t: \mathcal{V}\rightarrow \{1,2,\hdots,|\rho(\Phi_q)|\}$ be a random mapping between any node and its tree number in $\Phi_q$. (\eg, $t(i)=k$ if $i \in \mathcal{V}_k\in\pi(\Phi_q)$). By conditioning the root probability over a fixed partition $\mathcal{P}$, one obtains (see Proposition \mrw{2.3} in \cite{avena_two_2017}):  
\begin{equation}
 \forall i,j\in\mathcal{V} \quad \mathbb{P}(r_{\Phi_q}(i) = j | \pi(\Phi_q) = \mathcal{P}) = \frac{\mathbb{I}(j \in \mathcal{V}_{t(i)})}{|\mathcal{V}_{t(i)}|} 
\label{eq:condprobij}
\end{equation}    
where $\mathbb{I}$ is the indicator function (\ie, it outputs 1 if the input statement is true and 0 otherwise). In other words, given a fixed partition $\mathcal{P}$, the root probability within each subset $\mathcal{V}_k$ is uniform over the nodes in $\mathcal{V}_k$. ~\\

\noindent \textbf{Extending to non-constant $q$}. All of these properties are
adaptable to the case of $q$ varying over nodes (with some changes). The root process $\rho(\Phi_Q)$ is also a DPP. However, the associated marginal kernel becomes:
\begin{equation}
	\ma{K} = (\ma{L} + \ma{Q})^{-1}\ma{Q}  \quad \text{with} \quad \ma{Q} = \diag (q_1,\hdots,q_n). 
	\label{eq:generalizedK}
\end{equation} 
Notice that this kernel is not co-diagonalizable with the graph Laplacian $\ma{L}$. Thus, the expected number of roots $\mathbb{E}[|\rho(\Phi_q)|]$ is not writable in terms of $\lambda_i$'s, but it is still equal to $\tr(\ma{K})$. Similarly, the root probability $\mathbb{P}(r_{\Phi_Q}(i) = j)$ remains $\ma{K}_{i,j}$ whereas the conditional probability in \mrw{Eq.}~\eqref{eq:condprobij} becomes: 
\begin{equation}
\forall i,j\in\mathcal{V} \quad \mathbb{P}(r_{\Phi_Q}(i) = j | \pi(\Phi_Q) = \mathcal{P}) = \frac{q_j\mathbb{I}(j \in \mathcal{V}_{t(i)})}{\sum_{k\in\mathcal{V}_{t(i)}}q_k} .
\label{eq:condprobijvarq}
\end{equation}

\section{RSF based estimators}
\label{sec:estimators}
In this section, we present our main results. We first recall the graph Tikhonov
regularization and interpolation problems. 
Then, we describe the RSF-based methods to solve them. 
We also provide some theoretical analysis of the performance of the methods. Finally, we show how to tune hyperparameters for the proposed estimators.
~\\

\noindent \textbf{Graph Tikhonov regularization.} For a given graph $\mathcal{G} = (\mathcal{V},\mathcal{E},w)$ and measurements $\vec{y}=(y_1,y_2,\hdots,y_n)^\top$ on the  $|\mathcal{V}|=n$ vertices, the Tikhonov regularization of $\vec{y}$ reads:
\begin{equation}
\hat{\vec{x}} = \argmin_{\vec{z} \in \mathbb{R}^n} \mu ||\vec{y} - \vec{z}||^2 + \vec{z}^\top\ma{L}\vec{z}
\label{eq:tikhonov}
\end{equation} 
where $\ma{L} \in \mathbb{R}^{n\times n}$ is the graph Laplacian of
$\mathcal{G}$. The solution of this minimization problem is: 
\begin{equation}
	\hat{\vec{x}} = \ma{K}\vec{y} \text{ with }  \ma{K} = (\ma{L} + \mu\ma{I})^{-1}\mu\ma{I}.
	\label{eq:TRconstantmu}
\end{equation}  
Interestingly, the matrix in this solution also appears in \nt{Eq.}~\eqref{eq:rootkernel} as the marginal kernel of the root process. This
correspondence plays a significant role for the proposed methods by connecting RSFs
to the Tikhonov regularization problem. 

In some important cases, instead of $(\ma{L}+\mu\ma{I})^{-1}\mu\ma{I}\,\vec{y}$ , the generalized solution $(\ma{L}+\ma{Q})^{-1}\ma{Q}\,\vec{y}$ is required where $\ma{Q}$ is an entry-wise non-negative diagonal matrix. 
For example, if we write the Tikhonov regularization of \nt{Eq.}~\eqref{eq:tikhonov} with another graph Laplacian such as the random walk Laplacian $\ma{L}_{rw}=\ma{D}^{-1}\ma{L}$, then the solution reads $\hat{\vec{x}}=(\ma{L}+\ma{Q})^{-1}\ma{Q}\,\vec{y}$ where $\ma{Q}=\mu\ma{D}$.
Another example occurs when the noise variance is known to be non-constant over vertices, \ie~heteroscedastic noise. The measurements may be less reliable at some vertices compared to others, meaning that there are different noise variances $\sigma_1,\hdots,\sigma_n$. 
This implies that $q_1\hdots,q_n$ should be set proportional to $\frac{1}{\sigma_1},\hdots,\frac{1}{\sigma_n}$ in the estimation of $\hat{\vec{x}}$. This again corresponds to the generalized formulation.
~\\

\noindent \textbf{Graph interpolation.} Given a connected graph $\mathcal{G} = (\mathcal{V},\mathcal{E},w)$, a parameter $\mu\geq0$, and $\ell\subset\mathcal{V}$ a set of nodes where a signal $\vec{x}$ is known, \mrw{one way of defining the interpolated signal is~\cite{Pesenson2011}: }
\rwchng{\begin{equation}
	\begin{split}
		\hat{\vec{x}}=\argmin_{z\in\mathbb{R}^n}  \vec{z}&^\top(\ma{L}+\mu \ma{I})\vec{z}  \\ 
		\text{subject to }&   \forall i \in \ell, \text{ } z_i = x_i .
	\end{split}
	\label{eq:interpolation}
\end{equation}}
Define $u=\mathcal{V}\backslash\ell$ the set of nodes for which $\vec{x}$ is not known and write $\ma{L}$ in block form:
\[
\ma{L} = \begin{bmatrix}
	\ma{L}_{\ell|\ell} & \ma{L}_{\ell|u}\\
	\ma{L}_{u|\ell} & \ma{L}_{u|u}\\
\end{bmatrix}
\]
where $\ma{L}_{\ell|u}$ is the Laplacian reduced to its rows and columns indexed by $\ell$ and $u$, respectively.
The solution of \mrw{Eq.}~\eqref{eq:interpolation} reads: 
\begin{equation}
	\hat{\vec{x}} = \begin{cases}
		x_i& \quad \text{if } i \in \ell \\
		\left(-(\ma{L}_{u|u}+\mu \ma{I})^{-1}\,\ma{L}_{u|l}\,\vec{x}_\ell\right)_i& \quad \text{otherwise}\\
	\end{cases}
\label{eq:interpolsolution}
\end{equation}
where $\vec{x}_\ell\in\mathbb{R}^{|\ell|}$ is the signal $\vec{x}$ reduced to its entries in $\ell$. 
This solution can almost always\footnote{\label{footnote:mu_zero}$\ma{Q}$, as defined in the paragraph following \mrw{Eq.}~\eqref{eq:interpolscheme}, needs to be invertible for $\vec{y}$ to be well-defined. This is always the case if $\mu>0$. When $\mu=0$, it may not be the case. We will see in Section~\ref{sec:ssl} what can be done in this scenario.} be rewritten as: 
\begin{equation}
	\hat{\vec{x}}_u =\ma{K}\vec{y} \quad \text{with } \begin{cases}
	\ma{K}=(\ma{L}_{\mathcal{G}\setminus\ell}+\ma{Q})^{-1}\ma{Q}\\
	\vec{y}=-\ma{Q}^{-1}\ma{L}_{u|\ell}\vec{x}_\ell\\
	\end{cases}
	\label{eq:interpolscheme}
\end{equation}
where $\ma{L}_{\mathcal{G}\setminus\ell}$ is the Laplacian of the reduced graph obtained by removing the vertices (and the incident edges) in $\ell$, $\ma{Q}\in\mathbb{R}^{|u|\times|u|}$ is a diagonal matrix with $\ma{Q}_{i,i} = \mu+\sum_{j\in\ell}w(i,j)$. Similarly to graph TR, the RSF-based estimator for interpolation proposed in this paper draws upon the connection between Eqs.~\eqref{eq:interpolscheme} and~\eqref{eq:generalizedK}.~\\

\noindent \textbf{Parameter selection}. 
The solution to graph TR \mrw{Eq.~\eqref{eq:TRconstantmu}} tends to the constant vector (equal to the average of $\vec{y}$) as $\mu\rightarrow0$, and to $\vec{y}$ for $\mu \rightarrow \infty$, where it suffers from underfitting and overfitting,  respectively. 
In the interpolation problem, as $\mu\rightarrow \infty$, no prior information gets propagated through the other vertices, and $\hat{\vec{x}}_u$ tends to the zero vector.
The case of $\mu=0$ corresponds to solving the Dirichlet problem~\cite{Grady2003} which does not necessarily give the closest inference to the original signal $\vec{x}$. 
Due to these reasons, $\mu$ needs to be set at a value that gives the best approximation to the original signal. 
In both problems, choosing $\mu$ is a classical hyperparameter selection problem which can be approached in several ways for the proposed estimators. 

In the following, we first present the proposed estimators for approximating $\hat{\vec{x}}$ for a fixed value of $\mu$ in Section~\ref{sub:smoothing}. Then, we outline in Section~\ref{sub:trace} several methods that select an appropriate $\mu$ automatically. Combining the hyperparameter selection and the RSF based estimators forms an RSF-based  framework to approximate the solutions of graph Tikhonov regularization and graph interpolation. 

\subsection{RSF based estimation of $\hat{\vec{x}} = \ma{K}\vec{y}$}
\label{sub:smoothing}

We propose two novel Monte Carlo estimators to approximate $\hat{\vec{x}} = \ma{K}\vec{y}$ with $\ma{K} =(\ma{L}+\ma{Q})^{-1}\ma{Q}$. These estimators leverage the probability distribution of the root process on RSFs presented in \mrw{Eq.}~\eqref{eq:probij}. 
~\\

\begin{figure}
\centering
\scalebox{0.50}{
            \begin{tikzpicture}[
            every edge/.style = {draw=black,very thick},
             vrtx/.style args = {#1/#2}{%
                  circle, draw, thick, fill=white,
                  minimum size=5mm, label=#1:#2,scale =1},
                  every label/.append style={ font=\fontsize{10}{0}\selectfont},
                      baseline=(current bounding box.center)
                ]
            \node(A) [vrtx=center/1,fill=blue] at (-7, -1.5) {};
            \node(B) [vrtx=center/2,fill=green] at (-7.5, -2.5) {};
            \node(C) [vrtx=center/3,fill=yellow] at (-6.5,-2.5) {};
            \node(D) [vrtx=center/4,fill=red] at (-6,-3.5) {};
            \node(E) [vrtx=center/5,fill=pink] at (-9, -3.5) {};
            \node(F) [vrtx=center/6,fill=orange] at (-7.5, -3.5) {};
            \node(G) [vrtx=center/1,fill=blue] at (-6.5,-4.5) {};
            \node(H) [vrtx=center/3,fill=yellow] at (-8.5,-4.5) {};
                \path   (A) edge (B)
                (B) edge (C)
                (A) edge (C)
                (C) edge (D)
                (F) edge (D)
                (B) edge (F)
                (G) edge (D)
                (F) edge (H)
                (E) edge (H)
                (E) edge (F)
                (G) edge (F)
                (B) edge (E);
               	\hspace{-1.75cm}

                \draw[->,line width = 2] (-4,-3) to node[black,sloped,centered, above, text width=0.5cm,font=\fontsize{10}{10}\selectfont]{$\Phi_q$} (-3,-3);
            \end{tikzpicture}
		   \hspace{-1.75cm}
            \begin{tikzpicture}[
            every edge/.style = {draw=black,very thick},
             vrtx/.style args = {#1/#2}{%
                  circle, draw, thick, fill=white,
                  minimum size=5mm, label=#1:#2,scale =1},
                  every label/.append style={ font=\fontsize{10}{0}\selectfont},
                      baseline=(current bounding box.center)
                ]
            \node(A) [vrtx=center/1,fill=blue] at (-7, -1.5) {};
            \node(B) [vrtx=center/2,fill=green] at (-7.5, -2.5) {};
            \node(C) [vrtx=center/3,fill=yellow] at (-6.5,-2.5) {};
            \node(D) [vrtx=center/4,fill=red] at (-6,-3.5) {};
            \node(E) [vrtx=center/5,fill=pink] at (-9, -3.5) {};
            \node(F) [vrtx=center/6,fill=orange] at (-7.5, -3.5) {};
            \node(G) [vrtx=center/1,fill=blue] at (-6.5,-4.5) {};
            \node(H) [vrtx=center/3,fill=yellow] at (-8.5,-4.5) {};

                \draw[->]  (B) edge (A)
                (C) edge (A)
                (H) edge (E)
                (D) edge (C)
                (G) edge (F);
                \hspace{-1.5cm}

                \draw[->,line width = 2] (-4,-2) to node[black,sloped, anchor=center, above, text width=0.5cm,font=\fontsize{15}{15}\selectfont]{$\tilde{x}$} (-3,-1.5);
                \draw[->,line width = 2] (-4,-4) to node[black,sloped, anchor=center, above, text width=0.5cm,font=\fontsize{15}{15}\selectfont]{$\bar{x}$}  (-3,-4.5);
            \end{tikzpicture}
            \hspace{-1cm}
            \begin{tikzpicture}[
            every edge/.style = {draw=black,very thick},
             vrtx/.style args = {#1/#2}{%
                  circle, draw, thick, fill=white,
                  minimum size=5mm, label=#1:#2,scale =1},
                  every label/.append style={ font=\fontsize{10}{0}\selectfont},
                      baseline=(current bounding box.center)
                ]
            \node(A) [vrtx=center/1,fill=blue] at (-7, -1.5) {};
            \node(B) [vrtx=center/1,fill=blue] at (-7.5, -2.5) {};
            \node(C) [vrtx=center/1,fill=blue] at (-6.5,-2.5) {};
            \node(D) [vrtx=center/1,fill=blue] at (-6,-3.5) {};
            \node(E) [vrtx=center/5,fill=pink] at (-9, -3.5) {};
            \node(F) [vrtx=center/6,fill=orange] at (-7.5, -3.5) {};
            \node(G) [vrtx=center/6,fill=orange] at (-6.5,-4.5) {};
            \node(H) [vrtx=center/5,fill=pink] at (-8.5,-4.5) {};

                \draw[->]   
                (B) edge (A)
                (C) edge (A)
                (H) edge (E)
                (D) edge (C)
                (G) edge (F);                
               
              \hspace{-7.5cm}
              \node(A) [vrtx=center/2.5,fill={rgb:blue,1;green,1;yellow,1;red,1}] at (0.5, -5) {};
			  \node(B) [vrtx=center/2.5,fill={rgb:blue,1;green,1;yellow,1;red,1}] at (0, -6){};
			  \node(C) [vrtx=center/2.5,fill={rgb:blue,1;green,1;yellow,1;red,1}] at (1,-6) {};
			  \node(D) [vrtx=center/2.5,fill={rgb:blue,1;green,1;yellow,1;red,1}] at (1.5,-7) {};
			  \node(E) [vrtx=center/4,fill={rgb:pink,1;yellow,1}] at (-1.5, -7) {};
			  \node(F) [vrtx=center/3.5,fill={rgb:orange,1;blue,1}] at (0, -7) {};
			  \node(G) [vrtx=center/3.5,fill={rgb:orange,1;blue,1}] at (1,-8) {};
			  \node(H) [vrtx=center/4,fill={rgb:pink,1;yellow,1}] at (-1,-8) {};
			 \draw[->]   
       		 (B) edge (A)
        	 (C) edge (A)
       		 (H) edge (E)
       		 (D) edge (C)
       		 (G) edge (F);   
		\end{tikzpicture}
	}
    \caption{An illustration for the estimators where $q$ is constant over all nodes. In the left, the graph signal is interpreted by both colors and numbers. In the middle, a realization of $\Phi_q$, a forest, is illustrated. On this forest, the estimators $\tilde{\vec{x}}$ and $\bar{\vec{x}}$ are illustrated in top-right and bottom right, respectively.}
    \label{fig:estimators}
	\end{figure}
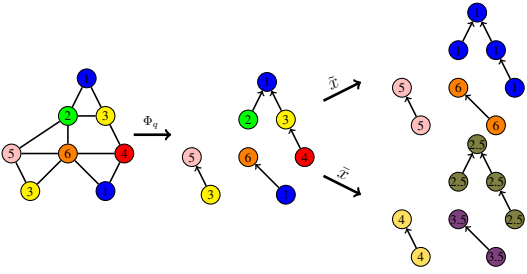

\noindent \textbf{The first estimator}, denoted by $\tilde{\vec{x}}$, is defined as follows: 
\begin{equation}
 \forall i \in \mathcal{V} \quad \tilde{x}(i) = y(r_{\Phi_Q}(i)) .
\end{equation} 
In practice, a realization of $\Phi_Q$ is considered. Then, in each tree, the measurement of the root is propagated through the nodes of the tree. (See top-right in Fig. \ref{fig:estimators}). 

\begin{proposition}
$\tilde{\vec{x}}$ is an unbiased estimator of $\hat{\vec{x}}$:
 \[\mathbb{E}\left[\tilde{\vec{x}}\right] = \hat{\vec{x}}. \]  
Moreover, the weighted expected error of $\tilde{\vec{x}}$ is:  
\[ \mathbb{E}\big(||\hat{\vec{x}} -\tilde{\vec{x}} ||^2_{\ma{Q}}\big) = \sum_{i\in\mathcal{V}} q_i\Var(\tilde{x}(i)) = \vec{y}^\top(\ma{Q}-\ma{K}^\top\ma{Q}\ma{K})\vec{y}\]
where $||\vec{x}||^2_{\ma{Q}} = \vec{x}^\top\ma{Q}\vec{x}$. 
\begin{proof}
For every node $i$, $\tilde{x}(i)$ is an unbiased estimator of $\hat{x}(i)$ thanks to the following:
\[\begin{split}\mathbb{E}\left[\tilde{x}(i)\right] =\mathbb{E}\left[y(r_{\Phi_Q}(i))\right] &=\sum_{j} \mathbb{P}(r_{\Phi_Q}(i)=j) y(j)\\
	&=\sum_{j} \ma{K}_{ij}y(j)=\bm{\delta}_i^\top\ma{K}\vec{y}=\hat{x}(i) \end{split} \]
where $\vec{\delta}_i$ is the Kronecker delta (\ie~${\delta}_i(i) = 1$ and 0 otherwise). This result is prominently due to the root probability of RSFs given in \mrw{Eq.}~\eqref{eq:probij}. Also, the variance of $\tilde{x}(i)$ reads:
\[\Var(\tilde{x}(i)) = \mathbb{E}\left[\tilde{x}(i)^2\right] - \mathbb{E}\left[\tilde{x}(i)\right] ^2 = \bm{\delta}_i^\top\ma{K}\vec{y}^{(2)}-(\bm{\delta}_i^\top\ma{K}\vec{y})^2.\]  	
where ${y}^{(2)}(k) = y(k)^2 $, $\forall k \in \mathcal{V}$. Then, the weighted sum reads:  
\begin{align*}
\begin{split}
\sum_{i\in\mathcal{V}}{q_i}\Var(\tilde{x}(i)) &=  \vec{1}^\top\ma{Q}\ma{K}\vec{y}^{(2)}-\vec{y}^\top\ma{K}^\top\ma{Q}\ma{K}\vec{y}
\end{split}
\end{align*}
where $\vec{1}$ denotes the all-ones vector. Note that $\vec{1}^\top\ma{Q}\ma{K} = \vec{1}^\top\ma{K}^\top\ma{Q}$. Moreover, $\vec{1}$ is a left eigenvector of $\ma{K}^\top$ with corresponding eigenvalue 1. Then, the first term becomes $\vec{1}^\top\ma{K}^\top\ma{Q}\vec{y}^{(2)} = \vec{1}^\top\ma{Q}\vec{y}^{(2)} = \vec{y}^\top\ma{Q}\vec{y}$, and, one obtains: 
\begin{equation}
	 \sum_{i\in\mathcal{V}}{q_i}\Var(\tilde{x}(i)) = \vec{y}^\top(\ma{Q} - \ma{K}^\top\ma{Q}\ma{K})\vec{y} .
\end{equation}
\end{proof}
\end{proposition}

\noindent \textbf{The second estimator}, denoted by $\bar{\vec{x}}$, is the
expectation of $\tilde{\vec{x}}$ conditioned on the partition induced by $\Phi_Q$: 
\begin{equation}
	\bar{x}(i) =\mathbb{E}[ \tilde{x}(i)|\pi(\Phi_Q)=\mathcal{P} ]  =  \frac{\sum\limits_{j\in\mathcal{V}_{t(i)}}y(j)q_j}{\sum\limits_{j\in\mathcal{V}_{t(i)}}q_j}.
\end{equation}
Due to the law of iterated expectations, this estimator is also unbiased, moreover it has a reduced variance compared to  $\tilde{x}(i)$ due to the law of total variance:
\[\Var(\tilde{x}(i)) = \mathbb{E}[\Var(\tilde{x}(i)|\pi(\Phi_Q)=\mathcal{P})] + \Var(\bar{x}(i))\]
which implies $\Var(\tilde{x}(i)) \geq \Var(\bar{x}(i))$. This idea of improving an estimator is often called Rao-Blackwellization~\cite{Blackwell1947,Rao1992}.  

In practice, we again take a realization of $\Phi_Q$ and consider the corresponding partition $\pi(\Phi_Q)$. Then, we compute the weighted average of the measurements in each subset of $\pi(\Phi_Q)$. Then, we finally propagate these averages in each subset (see Fig. \ref{fig:estimators}).   

\begin{proposition}
\label{prop:xbar}
$ \bar{\vec{x}}$  is an unbiased estimator of $\hat{\vec{x}}$:
	\[\mathbb{E}[\bar{\vec{x}}] = \hat{\vec{x}}\] 	
Moreover, the weighted expected error reads: 
	\[\mathbb{E}\left(||\hat{\vec{x}}-\bar{\vec{x}}||^2_{\ma{Q}}\right)=\sum_{i \in \mathcal{V}} {q_i}\Var(\bar{x}(i)) = \vec{y}^\top(\ma{Q}\ma{K}-\ma{K}^\top\ma{Q}\ma{K})\vec{y}.\]
\begin{proof}
Let $\ma{S} = \Big[\frac{q_j\mathbb{I}(j \in \mathcal{V}_{t(i)})}{\sum\limits_{k\in\mathcal{V}_{t(i)}}q_k}\Big]_{i,j} $ be a symmetric random matrix associated to the random partition $\pi(\Phi_Q)$. A simple matrix product shows that $\ma{S}^\top\ma{Q}\ma{S} = \ma{Q}\ma{S}$ by definition of $\ma{S}$. Moreover, $\bar{x}(i) = \bm{\delta}_i^\top\ma{S}\vec{y}$ and the expectation of $\ma{S}_{i,j}$ over all possible partitions derived from $\Phi_Q$ is:

\begin{equation}
\begin{split}
\mathbb{E}\left[\ma{S}_{i,j}\right] &=  \sum_{\mathcal{P} \in \pi(\mathcal{F})} \frac{q_j\mathbb{I}(j \in \mathcal{V}_{t(i)})}{\sum\limits_{k\in\mathcal{V}_{t(i)}}q_k}\mathbb{P}(\pi(\Phi_Q) = \mathcal{P}) \\
&=  \sum_{\mathcal{P} \in \pi(\mathcal{F})}\mathbb{P}(r_{\Phi_Q}(i) = j | \pi(\Phi_Q) = \mathcal{P})\mathbb{P}(\pi(\Phi_Q) = \mathcal{P}) \\ &=\mathbb{P}(r_{\Phi_Q}(i) = j ) = \ma{K}_{i,j}  .
\end{split}
\end{equation}
Similarly, the expectation of $\bar{x}(i)$ reads:
\begin{equation}
	\mathbb{E}[\bar{x}(i)] = \mathbb{E}[\bm{\delta}_i^\top\ma{S}\vec{y}] = \bm{\delta}_i^\top\mathbb{E}[\ma{S}]\vec{y} = \bm{\delta}_i^\top\ma{K}\vec{y} = \hat{x}(i).
\end{equation}
Thus, $\bar{\vec{x}}$ is unbiased. The expected error is also computed in a similar way: 

\begin{equation}
\begin{split}
\mathbb{E}\left(||\hat{\vec{x}}-\bar{\vec{x}}||^2_{\ma{Q}}\right) &= \sum_{i\in\mathcal{V}} {q_i} \Var(\bar{x}(i)) \\  
&=
\sum_{i\in\mathcal{V}} {q_i}\left(\mathbb{E}[(\bm{\delta}_i^\top\ma{S}\vec{y})^2] - \mathbb{E}[(\bm{\delta}_i^\top\ma{S}\vec{y})]^2\right)  \\ 
&= \sum_{i\in\mathcal{V}} \vec{y}^\top\mathbb{E}\left[{q_i}\ma{S}^\top\bm{\delta}_i\bm{\delta}_i^\top \ma{S}\right]\vec{y} -{q_i}\vec{y}^\top(\bm{\delta}_i^\top\mathbb{E}[\ma{S}])^2\vec{y} \\ 
&= \vec{y}^\top\mathbb{E}\left[\ma{S}^\top\ma{Q}\ma{S}\right]\vec{y} - \vec{y}^\top(\mathbb{E}[\ma{S}]^\top\ma{Q}\mathbb{E}[\ma{S}])\vec{y}	\\
&= \vec{y}^\top(\mathbb{E}\left[\ma{S}^\top\ma{Q}\ma{S}\right] - \ma{K}^\top\ma{Q}\ma{K})\vec{y}	.
\end{split}
\end{equation}
Finally, rewriting $\ma{S}^\top\ma{Q}\ma{S} = \ma{Q}\ma{S}$, one has: 
\[\mathbb{E}\left(||\hat{\vec{x}}-\bar{\vec{x}}||^2_{\ma{Q}}\right) = \vec{y}^\top(\mathbb{E}\left[\ma{Q}\ma{S}\right] - \ma{K}^\top\ma{Q}\ma{K})\vec{y} = \vec{y}^\top(\ma{Q}\ma{K} - \ma{K}^\top\ma{Q}\ma{K})\vec{y}.\]  
\end{proof}
\end{proposition}
\noindent \textbf{Sample Mean.} 
The sample mean of an unbiased Monte Carlo estimator over different realizations has a reduced variance, and so, gives a better estimator. Thus, in the rest, we use the sample means $\frac{1}{N}\sum_{k=1}^N\tilde{\vec{x}}_{\Phi^{(k)}_Q}$ and $\frac{1}{N}\sum_{k=1}^N\bar{\vec{x}}_{\Phi^{(k)}_Q} $ over $N$ forest realizations as the outputs of the RSF based methods.   
~\\

\noindent \textbf{A remark.} Reducing these results to the constant $q$ case ($\ma{Q}=q\ma{I}$), one recovers the preliminary results presented in~\cite{pilavci2020}. 
\subsection{Parameter selection for the RSF estimators}
\label{sub:trace}
The proposed estimators are efficient tools to approximate $\hat{\vec{x}}$ in both graph TR and interpolation problems for a fixed value of $\mu$. 
However, as usual in these problems, a difficult question is the tuning of the hyper-parameter: the choice of $\mu$ that yields the best performance. 
For linear smoothers such as the one we have at hand ($\hat{\vec{x}} =
\ma{K}\vec{y}$), many methods such as \rwchng{Akaike information criterion
  (AIC), Bayesian information criterion (BIC)}, Marlow's Cp, leave-one-out cross
validation (LOOCV), generalized cross validation (GCV) or Stein's unbiased risk
estimator (SURE) are readily available\footnote{\rwchng{Note that the theoretical framework used to justify these various
		criteria does not directly apply to graphs, because large-$n$ asymptotics are
		not immediately well-defined in graphs. One may appeal to random graph
		assumptions, as in \cite{keriven2020convergence}, but a detailed argument is
		beyond the scope of this paper.}} for this tuning step (for more details
and motivations, we refer the reader to \cite{hastie_elements_2005}). 

All of these methods need to compute a quantity called the effective number of parameters or the degree of freedom~\cite{hastie_elements_2005}, which equals $\tr(\ma{K})$ for linear smoothers of the form $\ma{K}\vec{y}$.
Computing exactly this trace requires the matrix inversion we wish to avoid from the start. 
A classical estimator of this quantity is Girard's estimator~\cite{girard1987algorithme} (also known as Hutchinson's estimator~\cite{Hutchinson1990}). 
We showed in~\cite{Barthelme2019} that RSFs can also be used to efficiently estimate $\tr(\ma{K})$. 
In this section, we build upon these preliminary results to show how the SURE and LOOCV methods can be adapted to the proposed estimators in order to select a good  value of $\mu$. 
Other methods are adaptable in a similar fashion.  
~\\

\noindent \textbf{Stein's Unbiased Risk Estimator}. 
Given independent noisy measurements $\vec{y}=\vec{x}+\bm{\epsilon} \in\mathbb{R}^n$ with a Gaussian noise $\epsilon_i\mrw{\sim}\mathcal{N}(0,\sigma^2)$, let $\theta(\vec{y})$ be an estimate for the unknown quantity $\vec{x}$. 
$\texttt{SURE}(\vec{y},\theta(\vec{y}))$ provides an unbiased estimate of the expected error $\mathbb{E}_{\bm{\epsilon}}[|| \theta(\vec{y}) - \vec{x} ||_2^2]$.
For the linear smoother ${\theta}(\vec{y}) = \ma{K}\vec{y}$ with $\ma{K}=(\ma{Q}+\ma{L})^{-1}\ma{Q}$, the generic formula of SURE in~\cite{Tibshirani} can be adapted as:
\begin{equation}
\texttt{SURE}(\vec{y},\theta(\vec{y})) = - n\sigma^2 +  ||\vec{y} - \theta(\vec{y})||_2^2 + 2\sigma^2\tr(\ma{K})
\label{eq:sureformula}
\end{equation}
where the degree of freedom term is replaced with $\tr(\ma{K})$. 
The theory behind relies on Stein's lemma on multivariate Gaussians~\cite{Stein1981}. 
Note that this method requires prior knowledge on the noise variance $\sigma^2$ and it outputs an unbiased estimation of the error. 
Then, this error needs to be evaluated for different values of $\mu$ and select the value yielding the smallest error. 
~\\

\noindent \textbf{SURE for RSF estimators.} 
Similar to $\hat{\vec{x}}$, the RSF estimators benefit from optimising the value
of $\mu$. For this purpose, SURE can be used. 
In the following derivations, we present the adapted SURE formula for $\tilde{\vec{x}}$ and $\bar{\vec{x}}$. 
Moreover, these derivations show that numerically computing this formula is trivial after sampling $N$ spanning forests. 

Consider two random matrices $\tilde{\ma{S}}=\Big[\mathbb{I}(r_{\Phi_Q}(i) = j) \Big]_{i,j}$ and $\bar{\ma{S}}= \Big[\frac{q_j\mathbb{I}(j \in \mathcal{V}_{t(i)})}{\sum\limits_{k\in\mathcal{V}_{t(i)}}q_k}\Big]_{i,j} $ (previously defined as $\ma{S}$ in the proof of Prop. \ref{prop:xbar}). With these definitions, notice that $\tilde{\vec{x}} = \tilde{\ma{S}}\vec{y}$ and $\bar{\vec{x}} = \bar{\ma{S}}\vec{y}$ . Moreover, the proposed estimators can be written in the form of linear smoothers: 
\begin{equation}
\begin{split}
\tilde{\theta} (\vec{y})=\frac{1}{N}\sum_{k=1}^N\tilde{\vec{x}}_{\Phi^{(k)}_Q} &=  \frac{1}{N}\sum_{k=1}^N\tilde{\ma{S}}^{(k)}\vec{y}, \\ 
\bar{\theta} (\vec{y})=\frac{1}{N}\sum_{k=1}^N\bar{\vec{x}}_{\Phi^{(k)}_Q} &=  \frac{1}{N}\sum_{k=1}^N\bar{\ma{S}}^{(k)}\vec{y}
\end{split}
\end{equation}
where superscript $(k)$ denotes $k$-th realization of $\tilde{\ma{S}}$ or $\bar{\ma{S}}$. 

Then, one can also evaluate the formula in \mrw{Eq.}~\eqref{eq:sureformula} for $\tilde{\theta}(\vec{y})$ and $\bar{\theta}(\vec{y})$. For instance, we get for $\tilde{\theta}(\vec{y})$: 
\begin{equation}
\texttt{SURE}(\vec{y},\tilde{\theta}(\vec{y})) = - n\sigma^2 +  ||\vec{y} - \tilde{\theta}(\vec{y})||_2^2 + 2\sigma^2\tr\left(\frac{1}{N}\sum_{k=1}^N\tilde{\ma{S}}^{(k)}\right).
\end{equation}
The residual error is trivial to compute after sampling $N$ spanning forests. Moreover, this is also the case for the degree of freedom
term. A closer look at the trace shows: 
\begin{align*}
\begin{split}
\tr\left(\frac{1}{N}\sum_{k=1}^N\tilde{\ma{S}}^{(k)}\right) = \tr\left(\frac{1}{N}\sum_{k=1}^N\bar{\ma{S}}^{(k)}\right) = \frac{1}{N}\sum_{k=1}^N|\rho({\Phi^{(k)}_Q})|. \\ 
\end{split}
\end{align*}
This result yields that the trace term can be replaced with the average number of roots in the computation of  $\texttt{SURE}(\vec{y},\tilde{\theta}(\vec{y})) $ or $\texttt{SURE}(\vec{y},\bar{\theta}(\vec{y}))$. Thus, the SURE scores of both estimators are trivial to numerically compute after sampling $N$ spanning forests. 

Note that neither $\texttt{SURE}(\vec{y},\tilde{\theta}(\vec{y})) $ nor $\texttt{SURE}(\vec{y},\bar{\theta}(\vec{y}))$ is an unbiased estimator for $\texttt{SURE}(\vec{y},{\theta}(\vec{y}))$. 
Moreover, the estimation errors read: 
\begin{equation}
	\begin{split}
		\mathbb{E}\left[\texttt{SURE}(\vec{y},\tilde{\theta}(\vec{y}))\right] - \texttt{SURE}(\vec{y},{\theta}(\vec{y})) &=\mrw{\sum_{i\in\mathcal{V}}} \Var(\tilde{\theta}(\vec{y})_i)\geq 0, \\ 	\mathbb{E}\left[\texttt{SURE}(\vec{y},\bar{\theta}(\vec{y}))\right] - \texttt{SURE}(\vec{y},{\theta}(\vec{y})) &= \mrw{\sum_{i\in\mathcal{V}}}\Var(\bar{\theta}(\vec{y})_i)\geq 0  .
	\end{split}
	\label{eq:sureupperbound}
\end{equation}
Thus, $\texttt{SURE}(\vec{y},\tilde{\theta}(\vec{y})) $ and
$\texttt{SURE}(\vec{y},\bar{\theta}(\vec{y}))$ are (with high probability) upper-bounds for $\texttt{SURE}(\vec{y},{\theta}(\vec{y}))$. For large graphs, in which computing $\texttt{SURE}(\vec{y},{\theta}(\vec{y}))$ is prohibitive, these upper bounds \rwchng{might also} be useful since they can be obtained cheaply. 
~\\

\noindent\textbf{Leave-One-Out cross validation}. 
LOOCV is a simple and popular method for hyperparameter selection. Unlike SURE,
it does not require for the noise variance to be known. 
Keeping the same notation as above, LOOCV is defined as: (See Chapter 5.5 in~\cite{hastie_elements_2005}): 
\[	\texttt{LOOCV}(\vec{y}_{\ell},\theta(\vec{y}_{\ell})) = \frac{1}{|\ell|}\sum_{i\in\ell} (\theta^{-i}(\vec{y}_{\ell})_i  - y_i)^2 
\]
where $\theta^{-i}(\vec{y}_{\ell})$ is the estimation without using the $i$-th
measurement.

This method leaves $y_i$ out at the estimation stage, and calculates the
corresponding prediction error. The overall score is the average error over the vertices in $\ell$. 
Note that this method needs to compute  $\theta^{-i}(\vec{y}_{\ell})$ for each
individual $i$ which in general costs $n$ times the cost of the original
estimator on the full data. 
Fortunately, this formula simplifies to the following for linear estimators in the form of $\theta(\vec{y})=\ma{K}\vec{y}$~\cite{hastie_elements_2005}: 
\begin{equation}
	\texttt{LOOCV}(\vec{y}_{\ell},\theta(\vec{y})) = \frac{1}{|\ell|}\sum_{i\in\ell} \left(\frac{\theta(\vec{y}_{\ell})_i  - y_i}{1 - \ma{K}_{i,i}}\right)^2
	\label{eq:LOOCVformula}
\end{equation}
which avoids re-computation.
~\\

\noindent \textbf{LOOCV for RSF estimators.} Similar to SURE, this score can be adapted for the RSF based estimators. For example, in case of $\tilde{\theta}(\vec{x}_{\ell})$, it becomes: 
\begin{equation}
	\texttt{LOOCV}(\vec{y}_{\ell},\tilde{\theta}(\vec{y}_{\ell})) = \frac{1}{|\ell|}\sum_{i\in\ell} \left(\frac{\tilde{\theta}(\vec{y}_{\ell})_i  - {y}_i}{1 - \frac{1}{N}\sum_{k=1}^N\tilde{\ma{S}}^{(k)}_{i,i}}\right)^2
	\label{eq:tildeLOOCVformula}
\end{equation}
and for $\bar\theta$, it can be derived in the same way. Notice that every element in this expression is numerically available after sampling $N$ spanning forests. Thus, this score can be easily computed for both estimators.

\section{RSF estimators for other graph problems}
\label{sec:graphotherprobs}
In this section, we explore a few graph problems in which the RSF based estimators presented can replace expensive exact computations.  

\subsection{Node Classification in semi-supervised learning}
\label{sec:ssl}
Consider a dataset consisting of elements one wishes to classify. 
In the semi-supervised learning context, the class label of a few elements are supposed to be known \emph{a priori}, 
along with a graph structure encoding some measure of affinity between the different elements: the larger the weight of the edge connecting two elements, the closer they are according to some application-dependent metric, the more likely these two elements belong to the same class. 
The goal is then to infer all the labels given this prior information. 

Among many options to solve this problem, label propagation\mrw{~\cite{zhu_2002learning,Zhu2005}} and generalized SSL framework~\cite{avrachenkov2012generalized} are two well-known baseline approaches. 
In this section, we deploy $\tilde{\vec{x}}$ and $\bar{\vec{x}}$ to approximate the solutions given by these approaches.
~\\

\noindent\textbf{Problem definition.} Let us denote the labeled vertices by $\ell\subset\mathcal{V}$ \rwchng{(typically $|\ell| \ll |\mathcal{V}|$)} and the unlabeled ones by $u=\mathcal{V}\backslash\ell$. Assume $C$ distinct label classes and define the following  encoding of the prior knowledge for the $c$-th class:
\begin{equation}
\forall i \in \mathcal{V},  \vec{y}_c(i) = \begin{cases}
1 \text{ if } i \text{ is known to belong to the } c\text{-th class} \\ 
0 \text{ otherwise.}
\end{cases}
\label{eq:labelencode}
\end{equation}
The matrix $\ma{Y}= [\vec{y}_1 | \hdots | \vec{y}_{C} ] \in \mathbb{R}^{n\times C}$ thus encodes the prior knowledge. 
Many approaches to SSL formulate the problem as follows. First, for each class
$c$, compute the so-called ``classification function'', defined as:
\begin{equation}
	\vec{f}_c = \argmin_{\vec{z}_c\in\mathbb{R}^n} \mu\sum_{i\in\mathcal{V}}q'_i(y_c(i)-z_c(i))^2 + \vec{z}_c^\top\ma{L}\vec{z}_c
	\label{eq:sslgenericformulation}
\end{equation}
where $\mu$ and $q'_i$'s are regularization parameters: $\mu$ sets the global
regularization level, and each $q'_i$ acts entry-wise (when $q'_i$ is high, the
corresponding entry in \rwchng{$\vec{f}_c$} is close to the measurement \rwchng{$\vec{y}_c(i)$}).
Eq.~\eqref{eq:sslgenericformulation} has the following explicit solution: 
\begin{equation}
  \vec{f}_c = (\ma{L}+\ma{Q})^{-1}\ma{Q}\,\vec{y}_c
\end{equation}
where $\ma{Q}=\diag(q_1,\hdots,q_n)$ and $q_i=\mu q'_i$.
Thus, each classification function $\vec{f}_c$ can be viewed as a smoothed version of the prior knowledge encoded in $\vec{y}_c$.  As such, if 
\rwchng{$\vec{f}_c(i)$} is large, it implies that labels of class $c$ are relatively dense
around node $i$.
The last step in these SSL algorithms is to assign each node $i$ to the class  \rwchng{$\argmax_{c}{\vec{f}}_{c}(i)$}.

Label propagation and generalized SSL framework are two algorithms that
adapt this solution in different ways. 
In particular, by using different set of $q'_i$'s in \mrw{ Eq.}~\eqref{eq:sslgenericformulation}, label propagation \rwchng{corresponds to solving} graph interpolation and the generalized SSL framework may be understood as \rwchng{solving} a graph TR. 
Thus, the RSF estimators can be used to
approximate the solution for both algorithms. In the following, we discuss the corresponding parameter settings for these algorithms along with their RSF versions. 
~\\

\noindent \textbf{Label Propagation.} 
The label propagation algorithm\mrw{~\cite{zhu_2002learning,Zhu2005}} solves the Dirichlet problem for each class $c$, that is:
\begin{equation}
\begin{split}
\forall i\in\mathcal{V},&\quad \ma{L}\vec{f}_c(i)= 0  \\
\text{s. t. }  \forall i \in \ell, & \quad \vec{f}_c(i) = \vec{y}_c(i)
\end{split}
\label{eq:labelproplinsys}
\end{equation}
which is equivalent to \mrw{Eq.}~\eqref{eq:interpolation} for $\mu=0$. Defining the classification matrix $\ma{F} = [\vec{f}_1|\hdots |\vec{f}_{C}] \in\mathbb{R}^{n\times C}$, one thus has:
\begin{equation}
\ma{F}_{i,c} = \begin{cases}
\ma{Y}_{i,c}, &\text{ if } i \in \ell  \\ 
(-(\mathsf{L}_{u|u})^{-1}\mathsf{L}_{u|\ell}\ma{Y}_{\ell|:})_{i,c}, &\text{ otherwise }   \\ 
\end{cases}
\label{eq:LPsolution}
\end{equation}
where $\ma{Y}_{\ell|:}$ is the matrix $\ma{Y}$ restricted to rows in $\ell$. Note in passing that $\ma{F}$ corresponds to a special set of functions for graphs called \emph{harmonic functions}. 
Besides being the solution of Dirichlet boundary problem, they have interesting connections with electrical networks and random walks\mrw{~\cite{Zhu2005}}.

\mrw{Zhu~\cite{Zhu2005}} provide a simple algorithm to compute $\ma{F}$ without computing the inverse matrix.  Starting from an arbitrary initial $\ma{F}^{(0)}$, at each iteration $k$, the algorithm updates $\ma{F}^{(k)}\leftarrow\ma{D}^{-1}\ma{W}\ma{F}^{(k-1)}$. The iteration is completed by setting the known labels $\ma{F}^{(k)}_{\ell|:}$ to $\ma{Y}_{\ell|:}$. They prove that the output of this iteration converges to $\ma{F}$ as $k\rightarrow\infty$ (see Section 2.3 in~\cite{Zhu2005}).

Here, we provide an RSF-based estimator to approximate $\ma{F}$. Two scenarii are possible. The first (unlikely) scenario is when any node in $u$ is connected to at least one node in $\ell$. In this case, one can rewrite \mrw{Eq.}~\eqref{eq:LPsolution} as:
\begin{align}
	\ma{F} = -\ma{K}\;\ma{Q}^{-1}\ma{L}_{u|\ell}\ma{Y} \quad \text{with } 
\ma{K}=(\ma{L}_{\mathcal{G}\setminus\ell}+\ma{Q})^{-1}\ma{Q}
\end{align}
where $\ma{L}_{\mathcal{G}\setminus\ell}$ is the Laplacian of the reduced graph obtained by removing the vertices (and the incident edges) in $\ell$, $\ma{Q}\in\mathbb{R}^{|u|\times|u|}$ is a diagonal matrix with $\ma{Q}_{i,i} = \sum_{j\in\ell}w(i,j)$. The condition of this first scenario ensures that $\ma{Q}$ is indeed invertible; and RSFs on the reduced graph $\mathcal{G}\setminus\ell$ can thus estimate the columns of $\ma{F}$. 

However, when there exists at least one node in $u$ that is not connected to $\ell$, $\ma{Q}$ is no longer invertible and another approach is needed. In this second scenario, 
the parameters are defined over all vertices and set to $q_i = \alpha>0$ for $i\in\ell$ and $q_i = 0$ for $i\in u$. The following proposition guarantees that as $\alpha\rightarrow \infty$, the RSF estimator $\tilde{\vec{x}}$ with this setting approximates the solution given in Eq~\eqref{eq:LPsolution}.

\begin{proposition}
\label{prop:labelprop}
Given the parameters $q_i = \alpha>0$ for $i\in\ell$ and $q_i = 0$ for $i\in u$, as well as the input vector $\vec{y}_c\in\mathbb{R}^{n}$ for the RSF estimator $\tilde{\vec{x}}$, the following is verified:
\[
\lim_{\alpha\rightarrow\infty}\mathbb{E}\left[ \tilde{\vec{x}}\right] = \vec{f}_c.
\]
\begin{proof}
See the supplementary material for the detailed proof. 
\end{proof} 
\end{proposition}
From the random forest point-of-view, setting $q_i$ to infinity for all nodes $i$ in $\ell$, and to $0$ otherwise, implies that all possible realizations of $\Phi_{Q}$ have exactly the same root set: $\ell$. Thus, when using the estimator $\tilde{\vec{x}}$, the measurements in $\ell$ are not altered and are simply propagated to other vertices via the sampled random trees. In addition, the estimator $\bar{\vec{x}}$ boils down to $\tilde{\vec{x}}$ in this very specific case.
~\\
 
\noindent \textbf{Generalized SSL framework.} The generalized SSL framework proposed in~\cite{avrachenkov2012generalized} defines the classification function as follows: 
\[
\vec{f}_{c} = \frac{\mu}{\mu+2}\left(\ma{I} - \frac{2}{\mu+2}\ma{D}^{-\eta} \ma{W}\ma{D}^{\eta-1}\right)^{-1}\vec{y}_c
\]
where $\mu>0$ is the regularization parameter and \rwchng{$\eta$ controls the normalization of the graph Laplacian}. This formula can also be written as: 
\[ \vec{f}_{c} = \ma{D}^{1-\eta }\ma{K} \ma{D}^{\eta -1}\vec{y}_c \text{ with } \ma{K} = (\ma{L} + \ma{Q})^{-1}\ma{Q}\]
where $\ma{Q} = \frac{\mu}{2}\ma{D}$. Notice that the cumbersome part in this formula is to compute $\ma{K} \ma{D}^{\eta -1}\vec{y}_c$ and it can be approximated by the proposed estimators on the input vector $\vec{y}= \ma{D}^{\eta -1}\vec{y}_c $. Then, $\vec{f}_{c}$ is obtained by left-multiplying the result by $ \ma{D}^{1-\eta }$.
~\\
 
Both solutions, label propagation and the generalized SSL, can be considered as two different versions of a more generic optimization problem. Label propagation puts a very high confidence on the prior.  $\tilde{\vec{x}}$, for example, only propagates the measurements of the labeled vertices (from the second scenario's perspective). Whereas, in generalized SSL, lower confidence over the prior information is assumed, and thus, the propagation of other measurements, which are all set to 0 in this encoding, is authorized. The success of both methods depends on the correctness of these assumptions on the data. Section \ref{sec:experiments} provides empirical comparisons on benchmark datasets.     
\subsection{Non-quadratic convex functions and Newton's method} 
\label{sec:newtonsmethod}
Consider the following generalized optimization problem: 
\begin{equation}
	\hat{\vec{x}} =\argmin_{z\in\mathbb{R}^n} \mu f(\vec{z}) + \frac{1}{2}\vec{z}^\top\ma{L}\vec{z} 
	\label{eq:nonquadop}
\end{equation}
where $f:\mathbb{R}^{n}\rightarrow\mathbb{R}$ is a generic twice-differentiable
function (\eg~previously $f(\vec{z}) =  || \vec{y} -
\vec{z}||^2_2$).
A common \nt{scenario} is when $f$ is a log-likelihood, i.e. $f(\vec{z}) = \sum_{i=1}^n \log p(y_i|z_i)$. This is used when the assumption that the
observations are Gaussian (given the signal) is inappropriate, for instance when
the observations are discrete. In such cases $f$
is not a quadratic function and there is typically no closed-form solution for \ref{eq:nonquadop}.

In these cases, iterative approaches are often deployed. 
One popular approach among them is Newton's method. Let $L(\vec{z})$ denote the loss function, Newton's method follows the following iteration scheme: 
\begin{equation}
	\vec{z}_{k+1} = \vec{z}_{k} - \alpha (\ma{H}L(\vec{z}_k))^{-1}\nabla L(\vec{z}_k) \text{ with } \alpha \in [0,1]\text{, }
\end{equation}    
$\ma{H}$ and $\nabla$ are the Hessian and gradient operators, respectively and $\vec{z}_{k}$ denotes the estimation at iteration $k$. Note that, by definition of the Hessian operator, this method requires twice-differentiability for the loss function. 

Given this scheme, the methods proposed here may become useful for approximating the inversion \nt{at each iteration}. We illustrate this usage in the following setup.

Assume an independent Poisson distribution for each likelihood at $i$:
  \[P(y_i|\lambda = z_i) = \frac{z_i^{y_i}\exp(-z_i)}{y_i!}\] 
where $\lambda$ is the distribution parameter. This assumption is often made in image processing applications to eliminate shot noise~\cite{Lebrun2012}. Also, consider the slightly modified loss function: 
\begin{equation}
	L'(\vec{t})= -\mu \sum_{i=1}^n \log P({y}_i|\lambda=\exp({t}_i)) + \frac{1}{2}\vec{t}^\top\ma{L}\vec{t}   
	\label{eq:newtonloss}
\end{equation} 
where $\exp(t_i)=z_i$. The gradient  $\nabla L'(\vec{t})$ reads: 
\[\nabla L'(\vec{t}) =\mu\exp(\vec{t})-\mu\vec{y} + \ma{L}\vec{t} 
\]
where $\exp$ operates entry-wise on vectors. Then, the Hessian matrix becomes:
 \[\ma{H}L'(\vec{t}) = \mu\diag(\exp(\vec{t}))  + \ma{L}. \]
With these two ingredients, \nt{the iterative scheme becomes:}
 \[
  \vec{t}_{k+1}= \vec{t}_k -\alpha \left[ \mu\diag(\exp(\vec{t}_k))  + \ma{L}\right] ^{-1}(\mu\exp(\vec{t}_k)-\mu\vec{y} + \ma{L}\vec{t}_k) \]The update term, which requires an inverse operation, can be approximated by our RSF estimators. This approximation is achieved by setting $\ma{Q} =\mu\diag(\exp(\vec{t}_k)) $ and the graph measurements $\vec{y}'$ to $\mu^{-1}\diag(\exp(-\vec{t}_k))(\mu\exp(\vec{t}_k)-\mu\vec{y} + \ma{L}\vec{t}_k)$. This particular case yields the following computation: 
\[ \begin{split}
\ma{K}\vec{y}' &= (\ma{Q}+\ma{L})^{-1}\ma{Q}\vec{y}' \\ 
					  &= [\mu\diag(\exp(\vec{t}_k)) + \ma{L}] ^{-1}(\mu\exp(\vec{t}_k)-\mu\vec{y} + \ma{L}\vec{t}_k)  
\end{split}\] 
which equals to the update in Newton's method. Thus, the RSF estimators can be easily used to compute each update step with a cheap cost.
 
In the classical Newton's method~\ie~$\alpha=1$, convergence of the result is not guaranteed. It might diverge or \nt{get} stuck in a loop depending on the closeness of the initial point $\vec{t}_0$ to the solution. 
Guessing a good initial point is not an easy task and may
require expensive computations in high dimensions. Instead, modifying
$\alpha$ is a more applicable option to ensure convergence. Thus, Newton's method is often combined with an additional step at each iteration in
which $\alpha$ is reset accordingly. Line search
algorithms~\cite{boyd_convex_2004} are simple and well-understood methods for
this purpose. At each iteration, if needed, they damp the applied update by
shrinking $\alpha$. These methods provide convergence,
however, they may require more iteration steps w.r.t. the pure Newton's method
with a good initial point. In our case, the updates are stochastic and exact
convergence cannot be expected. 

\subsection{$l_1$-Regularization and iteratively reweighted least squares}
\label{sec:irls}
As with the data fidelity term, many alternatives for the regularization term
are also available. Among them, $l_1$-regularization~\cite{tibshirani2011solution} is
often deployed to obtain sparser solutions. In~\cite{sharpnack2012sparsistency}, Sharpnack~\etal~adapts $l_1$-regularization for graphs as follows:
\begin{equation}
	\hat{\vec{x}} = \argmin_{\vec{z} \in \mathbb{R}^n} \mu ||\vec{y} - \vec{z}||^2_2 + 
	||\ma{B}\vec{z}||_1
	\label{eq:l1norm}
\end{equation}
where $\ma{B}$ is the edge incidence matrix and $||\ma{B}\vec{z}||_1 = \sum\limits_{(i,j) \in \mathcal{E}} w(i,j)^{1/2}|z_i-z_j|$.  
In contrast to the $l_2$ regularization, a closed form solution for \nt{the} $l_1$ case is not available and iterative optimization schemes \nt{are again used} to find the solution. \nt{The iterative reweighted least square \cite{Burrus1994} (IRLS) method is one such option and we show here how our RSF based estimators can help compute its numerical bottleneck}. 

Let $\ma{M} = \diag(\abs(\ma{B}\vec{z}))^{-1}$ where $\abs$ is the entry-wise absolute value operator. \nt{Note that}
\[\begin{split} ||\ma{B}\vec{z}||_1 = \abs(\vec{z}^\top\ma{B}^\top)\vec{1} 
=\vec{z}^\top\ma{B}^\top\ma{M}\ma{B}\vec{z} \end{split} \] 
where $\vec{1}\in\mathbb{R}^m$ is the all-ones vector. \nt{The problem of Eq.~\eqref{eq:l1norm} can thus be re-written as:}
\begin{equation}
		\hat{\vec{x}} = \argmin_{\vec{z} \in \mathbb{R}^n} \mu ||\vec{y} - \vec{z}||_2^2 + 
		\vec{z}^\top\ma{B}^\top\ma{M}\ma{B}\vec{z},
\end{equation}
\nt{and the IRLS iterative loop reads}:
\[\vec{z}_{k+1} = (\rwchng{2}\mu\ma{I} + \ma{B}^\top\ma{M_k}\ma{B})^{-1}\mu\,\vec{y} \text{ with } \ma{M}_k = \diag(\abs(\ma{B}\vec{z}_k))^{-1}.\]
A more detailed derivation and the convergence analysis of this scheme can be found in \cite{Burrus1994}. Notice that, by definition,  $\ma{B}^\top\ma{M_k}\ma{B}$ equals to a reweighted graph Laplacian $\ma{L}_k$. Then, computing the update at each iteration step immediately reduces to solving a graph Tikhonov regularization. Thus, $\tilde{\vec{x}}$ or $\bar{\vec{x}}$ can be used for estimating the update. ~\mrw{See the supplementary materials for an illustration.}

\begin{figure*}[!t]
	\centering
	\subfloat[Original Image $\vec{x}$]{\includegraphics[width=3.5cm]{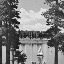}}%
	\hspace{1cm}	
	\subfloat[Noisy Measurments $\vec{y}$, \rwchng{PSNR = 14.0}]{\includegraphics[width=3.5cm]{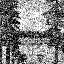}}%
	\hspace{1cm}	
	\subfloat[Exact solution $\hat{\vec{x}}$, \rwchng{PSNR = 19.2}]{\includegraphics[width=3.5cm]{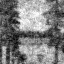}}%
	\hspace{1cm}	
	\subfloat[SURE scores vs $\mu$ \label{subfig:SURE}]{\includegraphics[width=4.25cm,height=3.5cm]{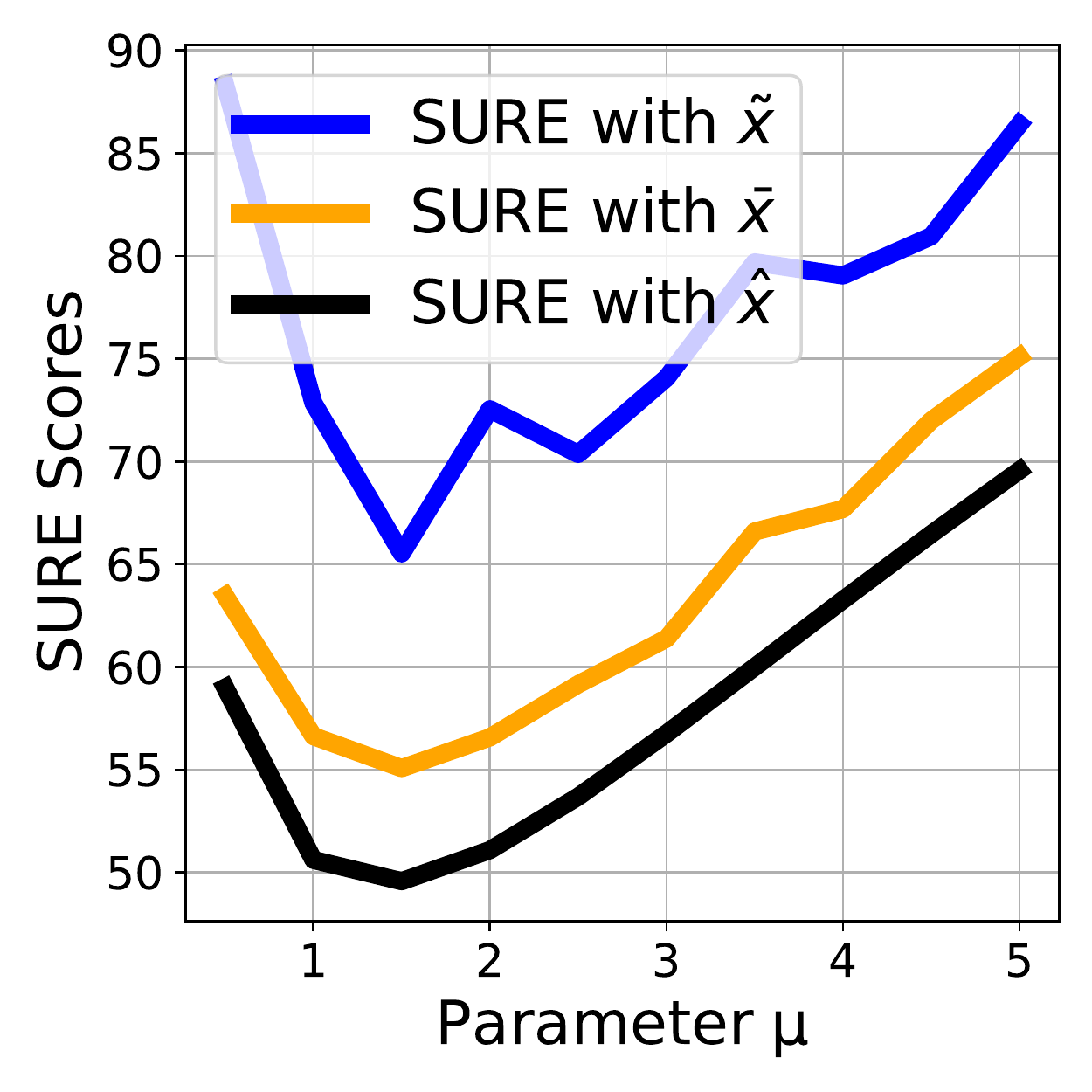}} \\
	\hspace{-0.75cm}
	\subfloat[$\tilde{\vec{x}}$ with $N = 1$, \rwchng{PSNR=12.2}]{\includegraphics[width=3.5cm]{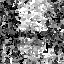}} 
	\hspace{1cm}	
	\subfloat[$\bar{\vec{x}}$ with $N = 1$, \rwchng{PSNR=16.1}]{\includegraphics[width=3.5cm]{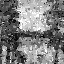}}%
	\hspace{1cm}	
	\subfloat[$\tilde{\vec{x}}$ with $N = 20$, \rwchng{PSNR = 18.3}]{\includegraphics[width=3.5cm]{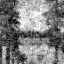}}%
	\hspace{1cm}	
	\subfloat[$\bar{\vec{x}}$ with $N = 20$, \rwchng{PSNR = 18.9}]{\includegraphics[width=3.5cm]{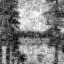}}%
	\caption{An image denoising experiment with additive Gaussian noise. a)~the
    original image. b)~a noisy version $\vec{y} = \vec{x} + \bm{\epsilon}$ with
    $\bm{\epsilon}\mrw{\sim}\mathcal{N}(\vec{0},\sigma )$ \rwchng{with $\sigma=0.2$}.
    c)~the exact \rwchng{solution} $\hat{\vec{x}}=\ma{K}\vec{y}$. Figure d) summarizes
    the SURE scores of $\hat{\vec{x}}$, $\tilde{\vec{x}}$ and $\bar{\vec{x}}$ for
    different $\mu$'s \rwchng{where $\mu \in \{0.5 ,1.0,1.5,\hdots 5.0\}$.} For
    each estimator, the value of $\mu$ that minimizes the SURE score is selected.
    \rwchng{For the computation of SURE scores, the noise variance is assumed to be known.} e-f)~the two RSF estimates $\tilde{\vec{x}}$ and $\bar{\vec{x}}$ based on only one sampled forest. g-h)~same as e-f) but averaged over $N=20$ sampled forests. \rwchng{Given the initial noisy image with PSNR$=14.0$, the exact solution produces a denoised image with PSNR$=19.2$ where the RSF estimates $\tilde{\vec{x}}$ and $\bar{\vec{x}}$ closely follows with PSNR=$18.3$ and $18.9$, respectively. 
} }
	\label{fig:gaussian_noise}
\end{figure*}

\section{Experiments}
\label{sec:experiments}
We provide in this section several illustrations and run time analysis of the proposed methods. First of all, in Section~\ref{subsec:image}, the RSF based methods are run on two image denoising setups. In these, we consider 
\begin{itemize}
\item An image corrupted by an additive Gaussian noise, for which we show its RSF-based Tikhonov regularization parameterized by SURE. 
\item An image corrupted by a Poisson noise, for which we show its version denoised by the RSF-based Newton's method coupled with line search. 
\end{itemize}
In both setups, the underlying graph is assumed to be a 2-dimensional (2D) grid
graph. \rwchng{We stress that these illustrations' main purpose is not to
  compete with the state-of-the-art image denoising methods but to provide
  visual examples for the proposed methods. Denoising performance can be
  improved in many ways, but it is not our goal here.}

Secondly, in Section~\ref{subsec:classif}, the SSL node classification problem  is considered on three benchmark datasets. We examine the classification performances of Tikhonov regularization, label propagation and the RSF versions of these algorithms. In these experiments, the parameter selection for the Tikhonov regularization is done by leave-one-out cross validation \rwchng{separately for each method}.

\rwchng{Finally, in Section~\ref{subsec:runtime}, we compare RSF estimators with the state-of-the-art methods,~\ie~Chebyshev polynomials and conjugate gradient method with and without preconditioning, in various graphs. Given a denoising setup, error vs. runtime plots are shown for all these algorithms.}

\subsection{Image denoising}
\label{subsec:image}
A 2D grid graph is a natural underlying structure for images: every pixel corresponds to a node and each pixel is connected to its four direct neighbors with equal weights. Other structures could be used (such as an 8-neighbour version of the grid graph) and performances will depend on the chosen structure. However, for the purpose of illustration, we will only consider the simplest 4-neighbour grid graph.

In the first setup, noisy (with additive Gaussian noise) measurements $\vec{y}$ of the original image $\vec{x}$ are given:
\[ \vec{y} = \vec{x} + \bm{\epsilon} \text{ with } \bm{\epsilon} \mrw{\sim} \mathcal{N}(\vec{0},\sigma^2\ma{I}) .\]
To recover $\vec{x}$, Tikhonov regularization is applied. Fig. \ref{fig:gaussian_noise} compares the exact result $\hat{\vec{x}}=\ma{K}\vec{y}= (\ma{L} + \mu\ma{I})^{-1}\mu\vec{y}$, to its forest-based approximations $\tilde{\vec{x}}$ and $\bar{\vec{x}}$. The SURE method to estimate the best value of $\mu$ is also illustrated for all three estimations of the original image $\vec{x}$ (and we remark in passing that they are consistent). 
\rwchng{The results confirm} that $\bar{\vec{x}}$ produces a better estimate for $\hat{\vec{x}}$ than $\tilde{\vec{x}}$. 
Also, in Fig.~\ref{subfig:SURE}, the scores computed for the RSF estimators are observed as upper bounds of the scores for $\hat{\vec{x}}$, as expected from the results in Eq~\eqref{eq:sureupperbound}.  
\nt{As observed in this illustration, and as expected by our theoretical analysis, $\bar{\vec{x}}$ always performs better than $\tilde{\vec{x}}$. In the following experiments and in order to avoid overloading the figures, we will omit the results obtained with $\tilde{\vec{x}}$.}

\begin{figure}
	\centering
	\subfloat[Original image $\vec{x}$]{\includegraphics[width=2.5cm]{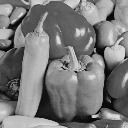}}%
	\hfil
	\subfloat[Noisy image $\vec{y}$, \rwchng{PSNR=17.1}]{\includegraphics[width=2.5cm]{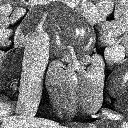}}
	\hspace{0cm}
	\subfloat[Loss function through iterations\label{subfig:newtoniterations}]{\includegraphics[width=3cm,height=2.5cm]{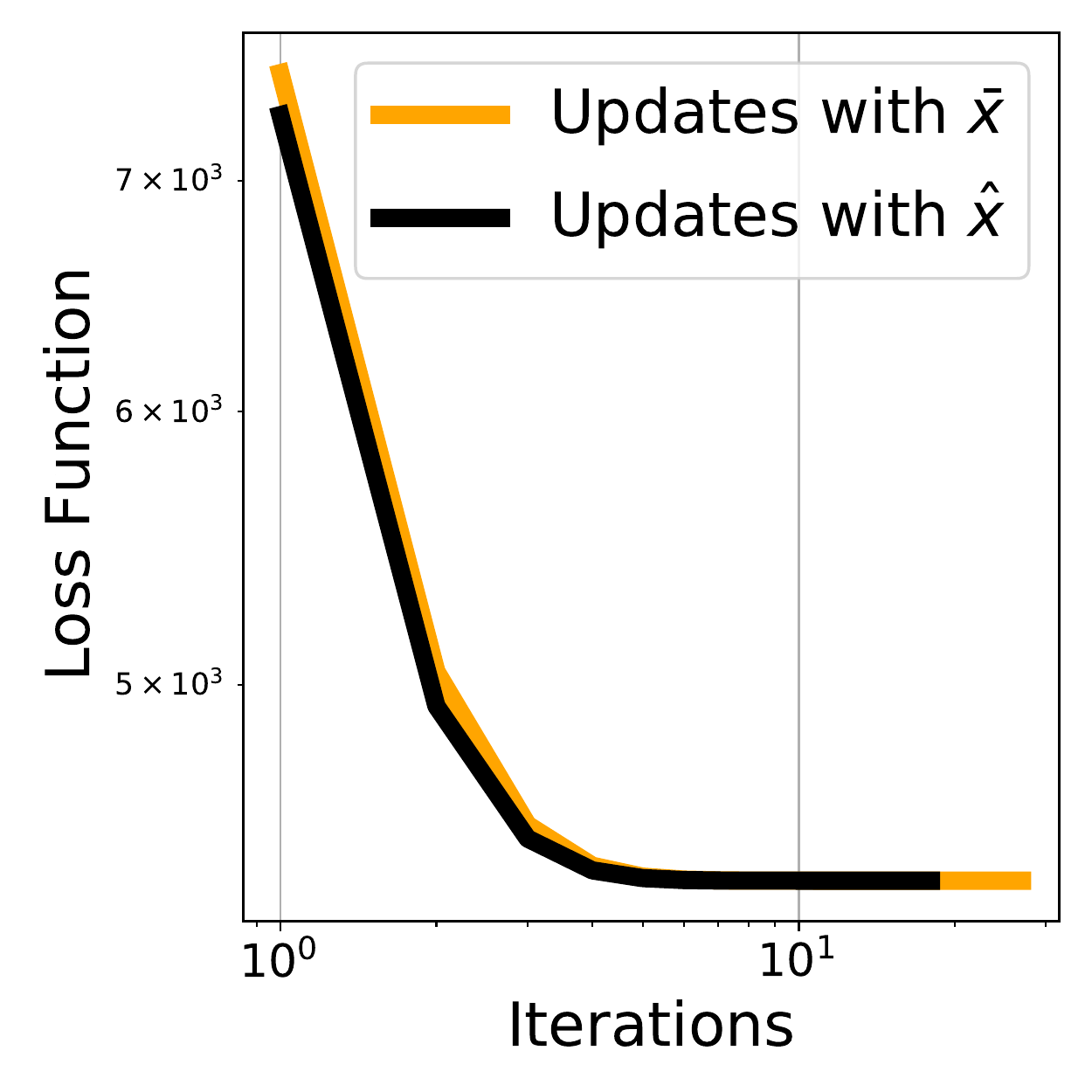}} \\ 
	\subfloat[Denoised image with $\hat{\vec{x}}$, \rwchng{PSNR=22.8}]{\includegraphics[width=2.5cm]{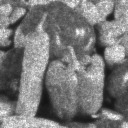}}
	\hfil
	\subfloat[Denoised image with $\bar{\vec{x}}$ with $N=40$, \rwchng{PSNR=22.8}]{\includegraphics[width=2.5cm]{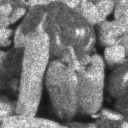}}

	\caption{
		An image denoising experiment with Poisson noise. a)~original image $\vec{x}$. b)~a noisy version $\vec{y} \mrw{\sim}\text{Poisson}(\vec{x})$. Newton's method is deployed to recover $\vec{x}$ by minimizing the loss function in \ref{eq:newtonloss}. For \mrw{two} update options, namely $\hat{\vec{x}}$ and $\bar{\vec{x}}$, Newton's method yields the results shown on the bottom line. Figure c) shows the loss function through the iterations for the \mrw{two} update options.}
	\label{fig:poisson_noise}
\end{figure}

In the second setup, each pixel value is assumed to be sampled from a Poisson distribution whose mean is the true value of the pixel: 
\[\vec{y} \mrw{\sim} \text{Poisson}(\vec{x}).\]   
To reconstruct $\vec{x}$, Newton's method is applied, as explained in Section \ref{sec:newtonsmethod}. The line search algorithm is used for picking the value of $\alpha$ at each iteration to ensure convergence. Both qualitative and quantitative results in Fig. \ref{fig:poisson_noise} show that 
\mrw{both $\hat{\vec{x}}$ and $\bar{\vec{x}}$ converge to the same solution. 
Fig.~\ref{subfig:newtoniterations} \nt{shows a similar decrease of the loss function for both $\hat{\vec{x}}$ and $\bar{\vec{x}}$, even though convergence is slightly faster for $\hat{\vec{x}}$.}}

\subsection{Node classification}
\label{subsec:classif}
In this illustration, we run our methods to solve the node classification problem discussed in Section \ref{sec:ssl}. For the generalized SSL framework, $\eta$ is set to $0$, and $\mu$ is set by RSF based cross-validation.

The experiments are done on three standard benchmark datasets, namely \rwchng{Cora
, Citeseer
and Pubmed}
\footnote{These datasets can be found in \url{https://linqs.soe.ucsc.edu/data} .}. In the first two, the underlying graphs are disconnected, thus we use the largest connected components. Also, in all datasets, the orientations of the edges are omitted to operate on undirected graphs. The general statistics of these datasets after the preprocessing are summarized in Table~\ref{tab:ssldatasetstat}. Note that in these three datasets, the class of each node is known. This will enable us to test the different SSL frameworks (a small arbitrary fraction of nodes will serve as pre-labeled nodes, and one tests whether or not this is sufficient to infer the class of all nodes). 
\begin{table}
\centering
\caption{SSL dataset statistics after preprocessing}
\begin{tabular}{c c c c }
\hline
Dataset & $\#$Nodes & $\#$Edges & $\#$Classes
\\
\hline 
\hline 
Citeseer & 2110 &  3668& 6 \\ 
\hline
Cora &2485  & 5069 & 7 \\ 
\hline
Pubmed & 19717 &44324  & 3   \\ 
\hline
\end{tabular}
\label{tab:ssldatasetstat}
\end{table}
More precisely, we use the following procedure: 
\begin{itemize}
\item $m$ vertices are selected at random per class as the labeled nodes,
\item the parameter $\mu$ is set by LOOCV separately for $\hat{\vec{x}}$ and $\bar{\vec{x}}$,
\item the classification functions $\vec{f}_c$ for each class $c$ are computed by the generalized SSL framework, label propagation and their RSF versions averaged over $N$ repetitions, 
\item for each vertex $i$, we assign \rwchng{$\argmax_{c}\ma{F}_{i,c}$} as its class and calculate the classification accuracy as the ratio of correctly predicted labels to the total number of predictions. 
\end{itemize}
\begin{figure}
	\includegraphics[width=9cm]{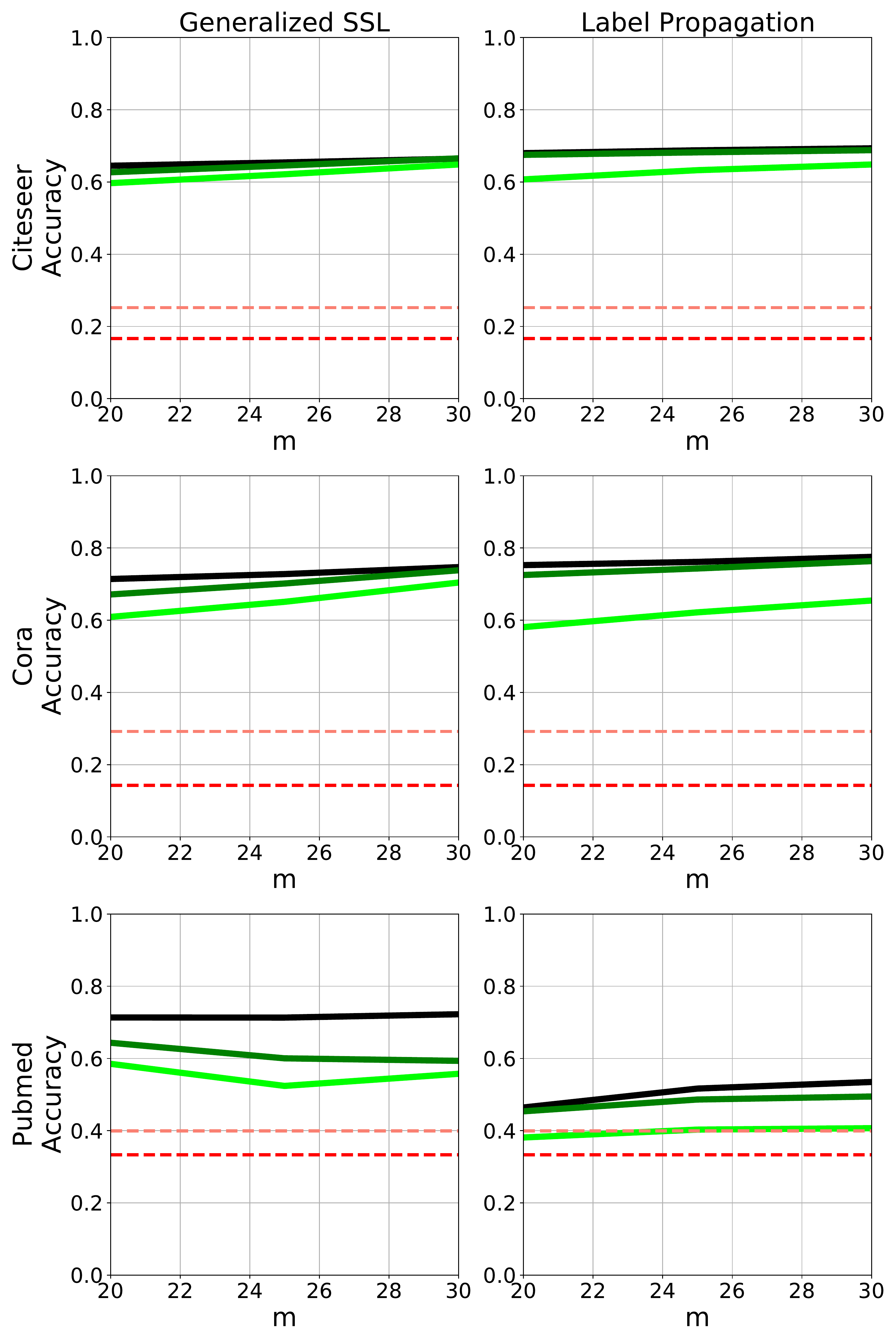}
	\includegraphics[width=9cm,height=0.75cm]{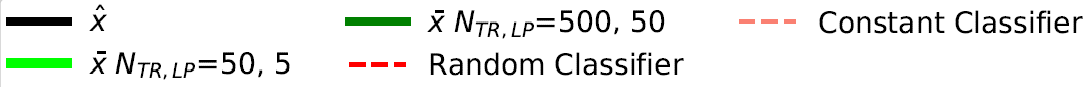}
	\caption{\rwchng{The classification accuracy of the generalized SSL, LP and their RSF variants are presented on the datasets Citeseer, Cora and Pubmed. The RSF methods for the generalized SSL are illustrated for $N=50,\text{ }500$ forest realizations, whereas, these numbers for LP are $N=5,\text{ }50$. 
	In these plots, the random classifier denotes the accuracy of inferring classes at random and the constant classifier is the accuracy of assigning the most occurring class to all unlabeled vertices. The results for Citeseer, Cora and Pubmed datasets are averaged over respectively 50 different set of labeled vertices.}}
	\label{fig:sslaccuracy}
\end{figure}

In Fig. \ref{fig:sslaccuracy}, the classification accuracy is reported as $m$ and $N$ vary. The results are averaged over 50 realizations of the $m$ labeled vertices for all datasets. 

For the first two datasets, Cora and Citeseer, 
$\bar{\vec{x}}$ has a comparable performance with the exact solution. However, in Pubmed, $\bar{\vec{x}}$ fails to perform as good as the $\hat{\vec{x}}$ for gSSL due to larger approximation errors in both the parameter selection and the estimation steps.  

The empirical results yield that the proposed methods need much less forest realizations to reach the exact solution of LP rather than the generalized SSL.  
However, sampling a forest for LP often takes more time if $n$ is large and $m$ is relatively small.
For example, in the Pubmed graph, for $m=20$, sampling a single forest for LP (resp. the generalized SSL) takes
$6.3\times10^{-2}$ (resp. $1.4\times10^{-3}$) seconds averaged over 100 repetitions in a single threaded run time of a laptop. Note that these figures strongly depend on $m$ and the given network. Thus, one needs to examine this trade-off with the given dataset to adjust the total run time.

\subsection{Run-time analysis}
	\label{subsec:runtime}
\rwchng{
In this section, we present an empirical comparison of the proposed
estimators with classical approaches, namely Chebyshev polynomial
approximation~\cite{Shuman2011}, and the conjugate gradient (CG) method (with
and without preconditioning)~\cite{Saad}. We run these algorithms on various
graphs and graph signals and report curves of performance versus run-time.  
} 

\rwchng{\subsubsection{Experimental setup}
In these experiments, we compute the result of the following graph smoothing operation via several methods:
\begin{equation}
	\hat{\vec{x}} = \ma{K}\vec{y} \text{ with } \ma{K} = (\ma{Q} + \ma{L})^{-1}\ma{Q}
	\label{eq:smoothing}
\end{equation}
where $\ma{Q}$ is set to $q\ma{I}$ for the
sake of simplicity in parameter tuning. Given a graph, the vector $\vec{y}$ is
assumed to contain noisy measurements of a \textit{$k$-bandlimited} signal
$\vec{x}$~\cite{puy_random_2016, di_lorenzo_sampling_2018}. The term ``bandlimited'' means that the signal belongs to a particular
subspace of harmonic functions and is defined more precisely below. 
The parameter $q$ is set to the value maximizing the denoising performance. In the following, these steps are further detailed. 
\\}

\noindent\rwchng{\textbf{Graph signal generation.} Consider $\ma{L} =
  \ma{U}^\top\ma{\Lambda}\ma{U}$, the eigendecomposition of $\ma{L}$ with the eigenvectors $\ma{U} = (\vec{u}_1|\vec{u}_2|\hdots|\vec{u}_n)$ and eigenvalues $\lambda_1=0\leq\lambda_2\leq\hdots\leq\lambda_n$ forming $\ma{\Lambda} = \diag(\lambda_1,\lambda_2,\hdots,\lambda_n)$. In GSP, $\ma{U} = (\vec{u}_1|\vec{u}_2|\hdots|\vec{u}_n)$ is considered to be a  Fourier basis for graph signals, and the $\lambda_i$'s are interpreted as generalized graph frequencies. For more on graph Fourier bases, we refer the reader to~\cite{Shuman2013, tremblay_design_2018}. In our experiments, we consider the denoising of a noisy version $\vec{y}$ of a bandlimited signal $\vec{x}$ defined as:
\begin{itemize}
	\item $\vec{x}$ is a $k$-bandlimited signal, \textit{i.e.}: 
	\[
	\vec{x} = \sum_{i = 1 }^k \alpha_i\vec{u}_i
	\]
	where the $\alpha_i$'s are the graph Fourier coefficients of $\vec{x}$. In other words, $\vec{x}$ is a low frequency graph signal, and is thus smooth on the given graph. 
	\item The noise is Gaussian: $\vec{y} = \vec{x} + \bm{\epsilon}$ with $\epsilon_i \sim \mathcal{N}(0,\sigma^2)$.
\end{itemize}
In the simulations, we randomly generate different realizations of $\vec{y}$ as follows:
\begin{enumerate}
	\item Sample $k=5$ Fourier coefficients $\alpha_i$'s from a Gaussian distribution $\mathcal{N}(0,1)$ and obtain a realization of $\vec{x}$.
	\item Normalize it to unit norm~\ie~$||\vec{x}||_2 =1$.
	\item Draw $n$ iid values from $\mathcal{N}(0,\sigma^2)$ to create the noise vector $\bm{\epsilon}$.  To obtain a given signal-to-noise ratio (the SNR is fixed to 2), the variance of the noise is set to $\sigma^2 = (n\times\text{SNR})^{-1}=1/2n$. 	
	\item Pass the noisy signal $\vec{y} = \vec{x} + \bm{\epsilon}$ to the algorithms. 
\end{enumerate}}
\bigskip
\noindent\rwchng{\textbf{Performance metrics.} Let $\vec{x}^\ast$ be the generic output
of the tested algorithms. In the experiments, we measure their performance with
respect to two metrics. The first one is the approximation error $ || \vec{\hat{x}} - \vec{x}^\ast||_2$ which evaluates the quality of the approximation to $\vec{\hat{x}} = \ma{K}\vec{y}$ achieved by the algorithms. The second is the reconstruction error $|| \vec{x} - \vec{x}^\ast||_2$ which measures the denoising performance of the output. 
\\}

\noindent\rwchng{\textbf{Parameter tuning.} The parameter $q$ is always set to the value yielding the best denoising performance (that is, to the value of $q$ minimizing $ || \vec{x} - \vec{\hat{x}}||_2$).
\\}

\noindent \rwchng{\textbf{Graphs.} We experimented on many graphs with varying structures and density. Table~\ref{tab:graphs} summarizes all graphs used in the experiments. 
\\}
\begin{table*}[!t]
	\centering
	\caption{Graphs considered in the run-time experiments}
	\begin{tabular}{p{0.15\textwidth}p{0.15\textwidth}p{0.15\textwidth}p{0.45\textwidth}}
		\hline
		Graph & $n$ & $m$ & description \\
		\hline 
		\hline 
		Grid & 10000  &  20000& 100x100 grid with periodic boundary conditions. \\ 
		\hline
		Barabasi-Albert & 10000 & 19996 & Randomly generated by the Barabasi-Albert model~\cite{barabasi1999emergence}. \\ 
		\hline
		Erdos-Renyi & 10000 & 49877& A random Erdos-Renyi graph with average degree around 10.   \\ 
		\hline
		K-regular & 10000 & 50000 & A random graph where each node has exactly 10 neighbors. \\ 
		\hline
		Euclidean & 10000 & 112613 &  (K=20)-nearest neighbor graph of $n$ randomly drawn points in $\mathbb{R}^3$.{ Whenever a node $i$ is a K-nearest neighbor of a node $j$, both edges $(i,j)$ and $(j,i)$ are added to symmetrize the graph. }  \\ 
		\hline
		Bunny & 2503 & 65490 &  ($\epsilon=0.2$)-Nearest neighbor graph of Stanford's bunny point cloud dataset~\cite{turk1994zippered}. Taken from the PyGSP toolbox~\cite{defferrard2017pygsp}. {Whenever a point is close to another one in a distance less than $\epsilon$, an edge between them is added.} \\ 
		\hline
		
		Citeseer &2110  &  3668& -\\ 
		\hline
		Cora &2485  & 5069 &  - \\ 
		\hline
	\end{tabular}
	\label{tab:graphs}
\end{table*}

\begin{figure*}
	\centering

	\fbox{\includegraphics[height=4cm,width=8.5cm]{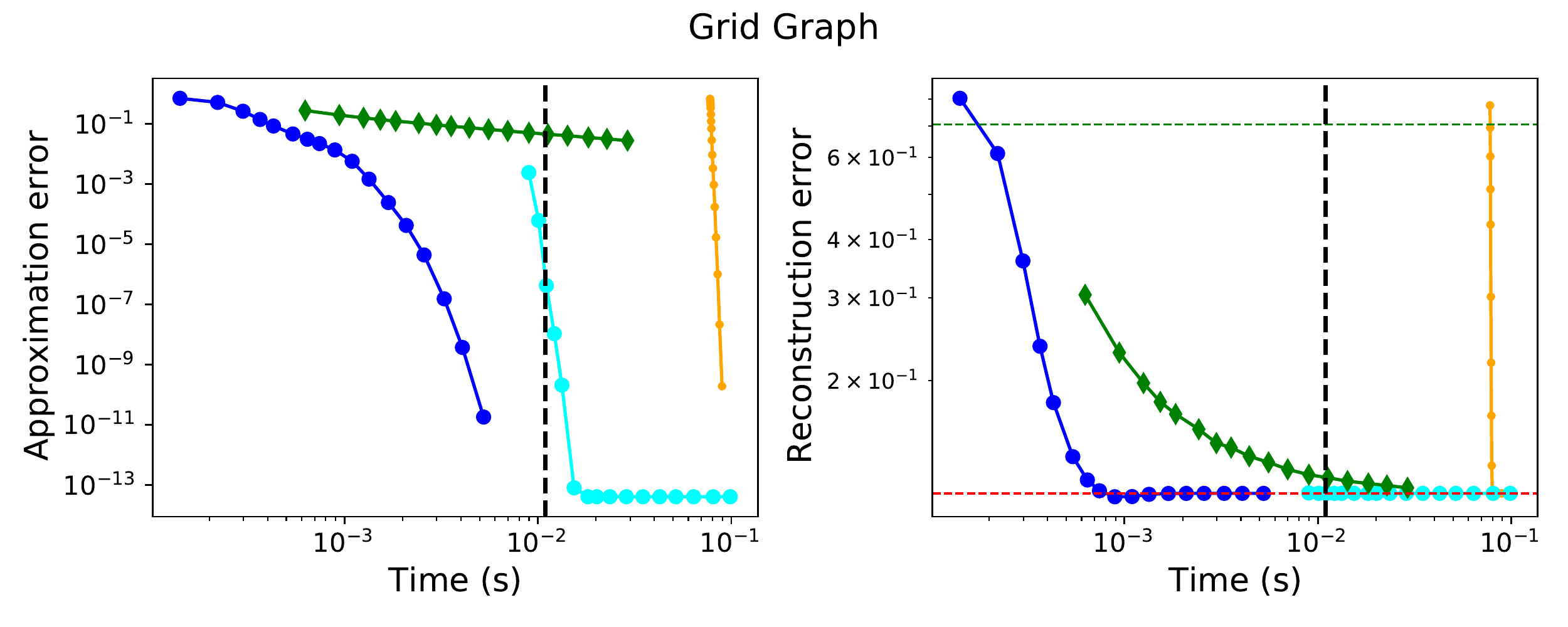}}
	\fbox{\includegraphics[height=4cm,width=8.5cm]{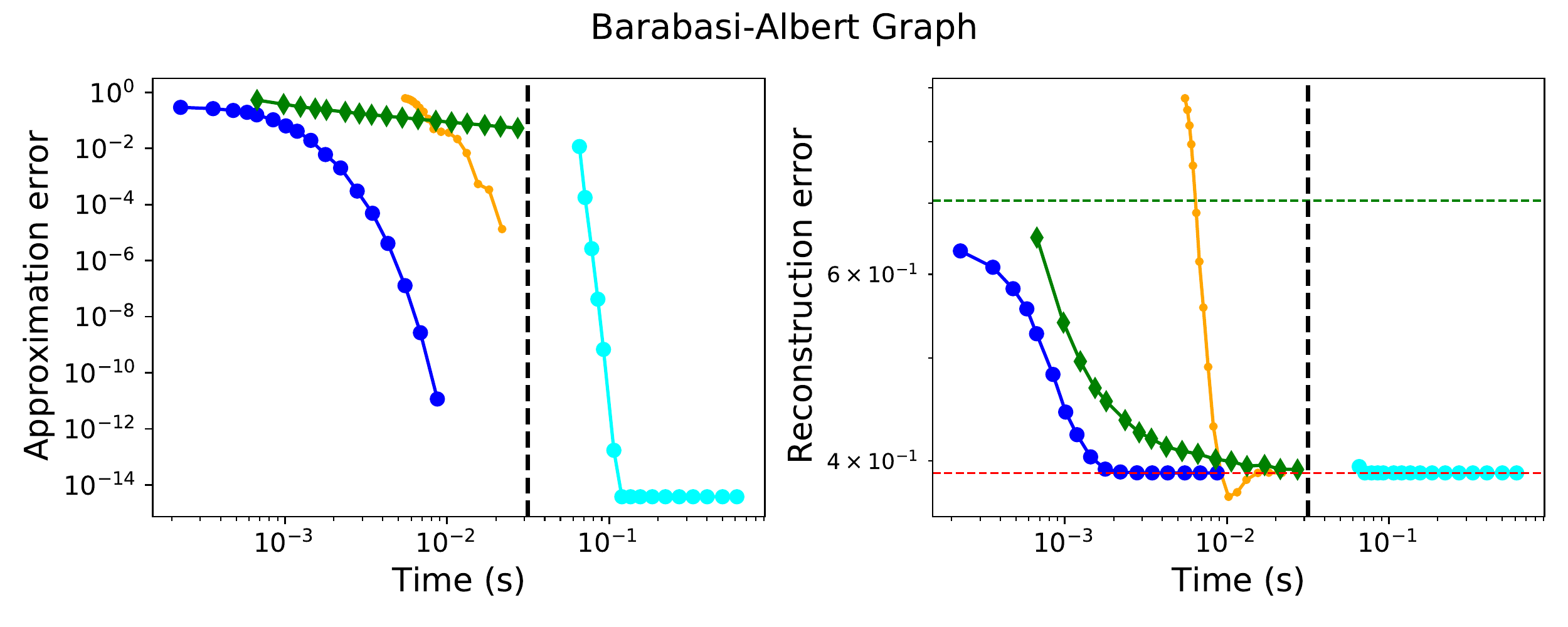}}
	
	\fbox{\includegraphics[height=4cm,width=8.5cm]{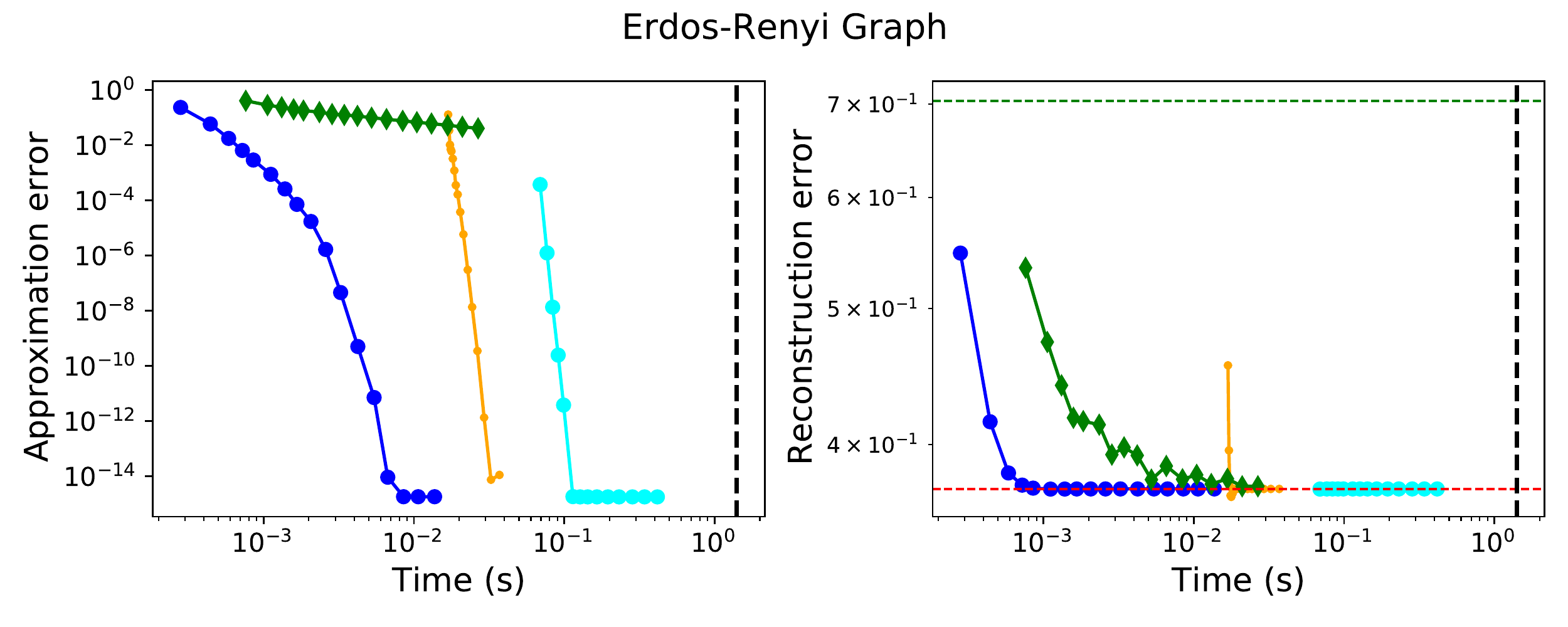}}
	\fbox{\includegraphics[height=4cm,width=8.5cm]{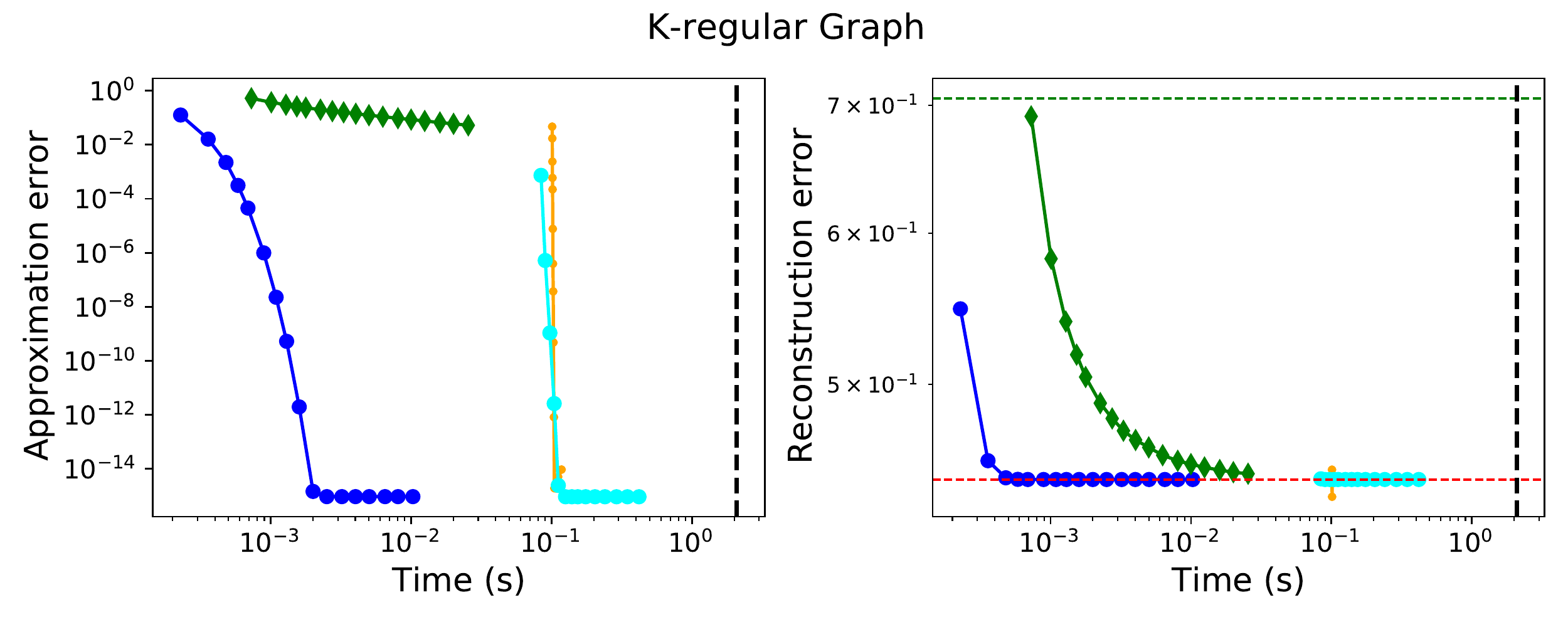}}
	
	\fbox{\includegraphics[height=4cm,width=8.5cm]{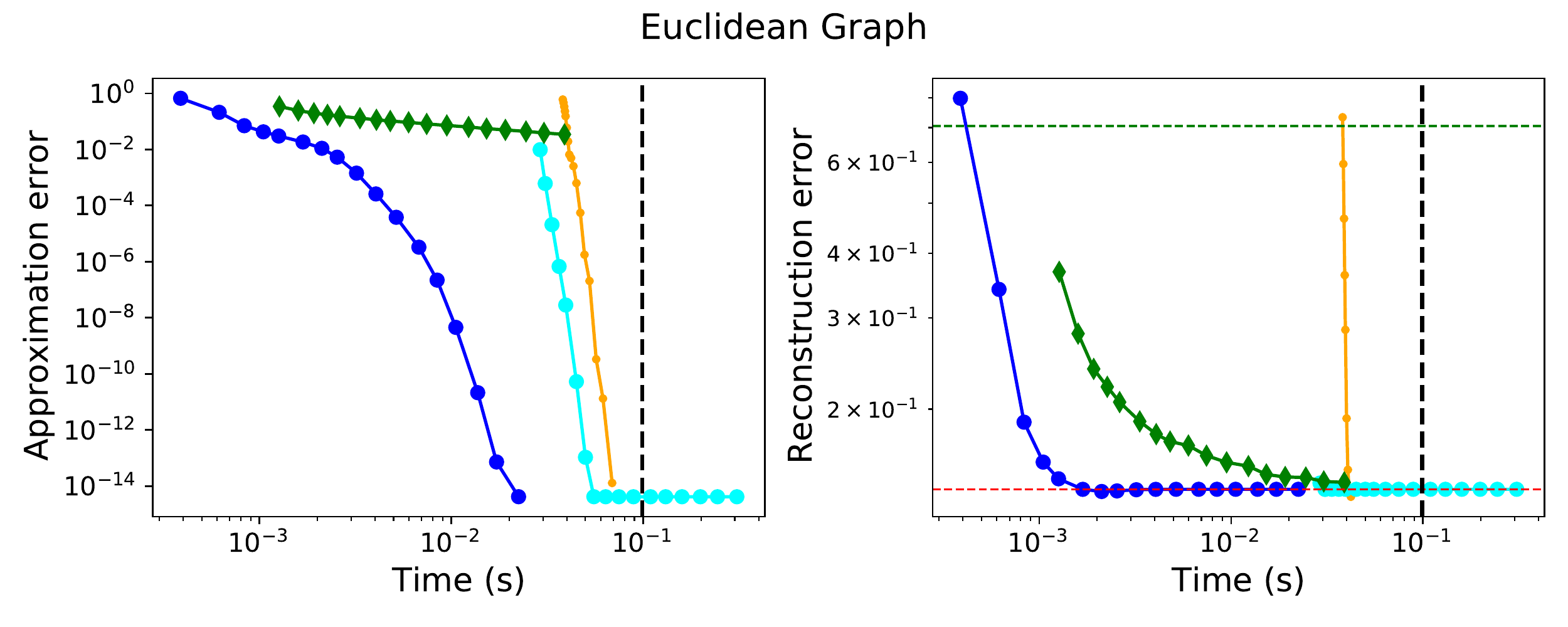}}
	\fbox{\includegraphics[height=4cm,width=8.5cm]{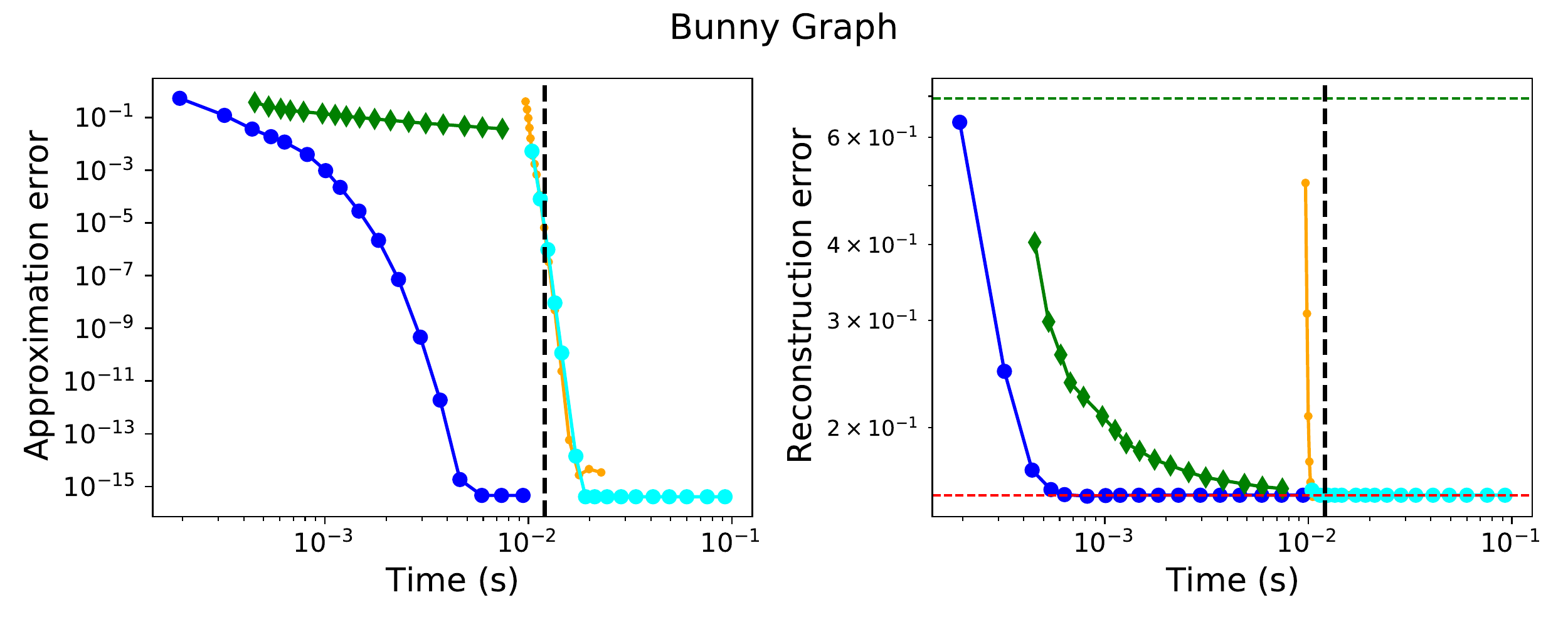}}
	
	\fbox{\includegraphics[height=4cm,width=8.5cm]{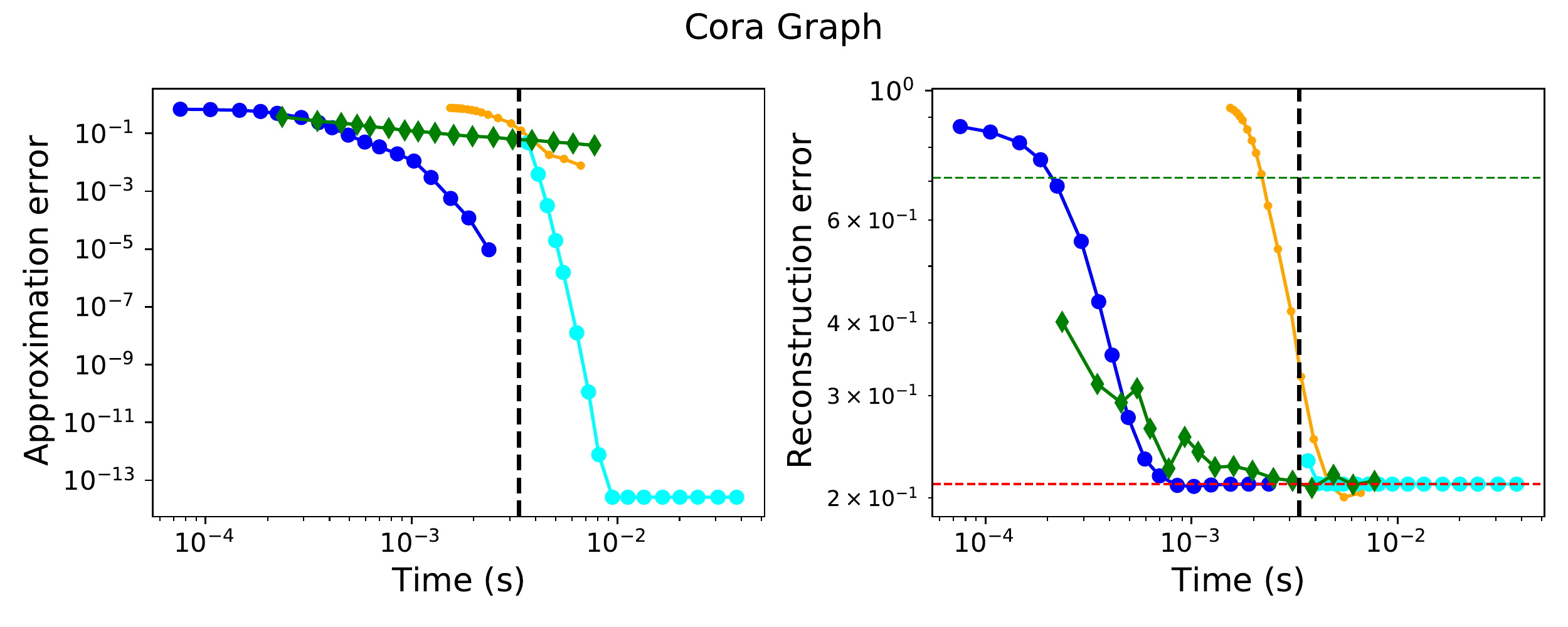}}
	\fbox{\includegraphics[height=4cm,width=8.5cm]{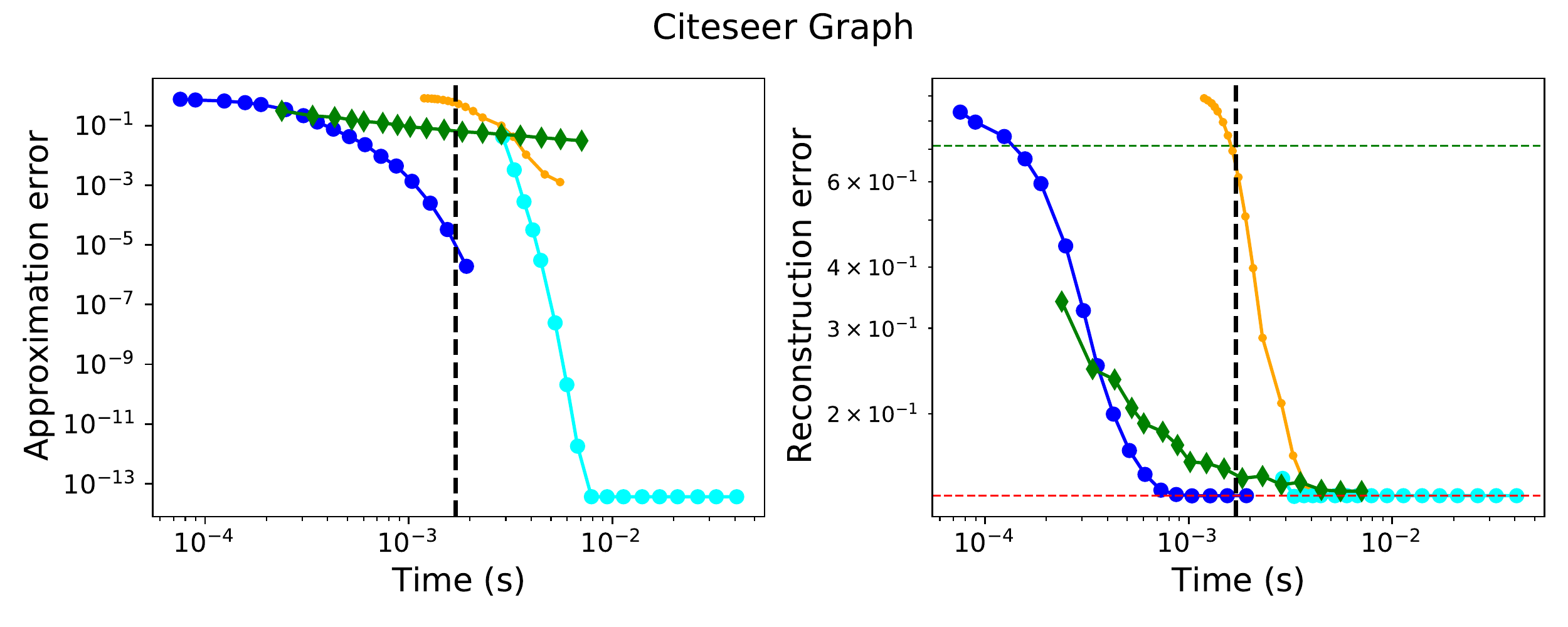}}
	
	\includegraphics[height=1.0cm,width=8cm]{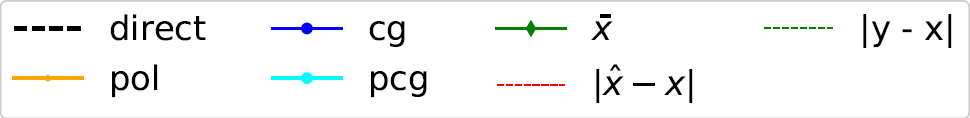}
	
	\caption{Approximation and Reconstruction error vs run-time for CG (dark blue), PCG (cyan), Chebyshev polynomial approximations (orange) and the RSF estimator $\vec{\bar{x}}$ (dark green) over various graphs. The vertical dashed line corresponds to the time to calculate the exact solution via the backslash operation of Julia (Sparse decomposition). The horizontal dashed red line indicate the reconstruction error of the exact solution $||\vec{x} - \hat{\vec{x}}||_2$. Finally, the horizontal green dashed line is the reconstruction error of the noisy signal $\vec{y}$. In all graphs and methods, we observe that the best denoising performance is reached many iterations before reaching the best approximation error. }
	\label{fig:results}
\end{figure*}

\noindent \rwchng{\textbf{Experimental procedure. } For a given graph, we follow this simple procedure: 
\begin{enumerate}
	\item Generate $\vec{y}$ as described earlier. 
	\item Find the value of $q$ that minimizes $||\vec{x} -\vec{\hat{x}} ||$ by a grid search. 
	\item Run the algorithms and measure the run-time, approximation error and reconstruction error. 
\end{enumerate}   
This procedure is repeated for different runs of the algorithms and realizations of $\vec{y}$, and the results are averaged. All simulations are done in a single thread of a laptop. Fig.~\ref{fig:results} summarizes the results. }

\rwchng{\subsubsection{Results}
The algorithms compared 
are: the RSF based estimators, Chebyshev polynomial approximation, conjugate
gradient method with and without preconditioning\footnote{Julia implementation
	in \url{https://julialinearalgebra.github.io/IterativeSolvers.jl/dev/}}. The
preconditioner for CG is the Algebraic Multigrid (AMG)\footnote{Julia implementation
	in \url{https://github.com/JuliaLinearAlgebra/AlgebraicMultigrid.jl}. The package's default parameters are used.} method~\cite{xu2017algebraic}. 
All of these algorithms admit an iteration parameter determining their performance:
the larger this iteration parameter, the better the approximation but the longer the computation time. These parameters are: 
\begin{enumerate}
	\item The number of forests for $\vec{\bar{x}}$ 
	\item The polynomial's degree in Chebyshev approximation
	\item The number of iterations for CG and preconditioned CG \mrw{(PCG)}.
\end{enumerate}
For each algorithm and each graph, this iteration parameter is swept through 17 logarithmically spaced values between 1 and 100: $\{1,2,\hdots,62,78,100\}$, corresponding to the 17 plotted markers forming each curve. In each curve, the marker at the top left-hand corner thus corresponds to $1$ iteration, whereas the marker at the bottom right-hand corner corresponds to $100$ iterations.
\mrw{The samples for error and time measurements are collected from separate runs}. The error results are averaged over 20 realizations of $\vec{y}$. The time
measurements are averaged over 100 runs for each realization of $\vec{y}$. For
the graphs generated from random graph models, such as Erdos-Renyi or
Barabasi-Albert, only a single realization of the graph is used in these results.} 

\rwchng{PCG and Chebyshev polynomial approximation are two algorithms that
  require some preprocessing. In PCG, this preprocessing is simply the calculation of
  the preconditioner. In the polynomial approximation method, one needs to provide the spectrum interval $[0,b]$ over which the approximation should be computed (the interval starts at $0$ since the smallest eigenvalue of $\ma{L}$ is necessarily null). For the polynomial approximation to work, $b$ should be an upper bound of $\lambda_n$, $\ma{L}$'s largest eigenvalue. A popular option\footnote{Another option is to set $b$ to the trivial upper bound equal to $2d_{\rm max}$ where $d_{\rm max}$ is the largest degree (this upper bound is easily obtained via Gershgorin's circle theorem). The preprocessing cost is in this case null but this upper bound has the drawback of being very crude for some graphs: a good polynomial approximation constrained on such a large interval will require an unnecessarily large polynomial order --inducing a longer computation time. } in the GSP literature, and the option we choose in this paper, is straightforward: compute $\lambda_n$ and set $b$ to $\lambda_n$. This ensures that the polynomial order required for a good approximation on $[0,b]$ is the lowest possible, thus minimizing the number of matrix-vector computations. However, the preprocessing step of computing $\lambda_n$, even via efficient Krylov-based routines, is not free and is in fact a significant part of the cost on the graph examples considered here. 
  In Fig.~\ref{fig:results}, we report the total time (preprocessing
  and algorithm run) for these algorithms. 
  Both algorithms have very steep convergence but their preprocessing costs make them stand out compared to the other methods: a more favorable context for them would be cases where one needs to filter \emph{many} graph signals, thus making their overhead cost worthwhile.
 }

\rwchng{Regarding the performance of $\vec{\bar{x}}$, we observe that
  \mrw{it follows} the usual Monte Carlo convergence regime
  (\ie, $\mathcal{O}(N^{-\frac{1}{2}})$, thus linear on log-scale) as expected.
  Due to this inherent weakness, RSF methods cannot compete with the state-of-the-art if the goal is to obtain high precision in the approximation of $\hat{\vec{x}}$. However, as the results also show, it is often not necessary to have very high precision in approximation to get a good reconstruction performance (no point
computing with high accuracy an inaccurate quantity). See
\cite{bottou2012stochastic} for a detailed discussion of this point in the
context of Stochastic Gradient Descent. When looking at reconstruction errors instead of approximation errors, RSF-based estimators are comparable to state-of-the-art methods (results depend on the graph at hand).}

\rwchng{Note that the current implementation of the RSF methods fail to benefit from compiler and hardware optimizations as much as the other deterministic methods do.
The deterministic methods use the matrix $\ma{L}$ only in the vector-matrix products.
In sparse implementations, these operations are well-optimized in terms of memory-access to the entries of $\ma{L}$. 
On the other hand, the current implementation of RSF methods does not have this
advantage. To sample a forest, the entries of $\ma{L}$ are
accessed randomly, which incurs additional cost as random parts of the matrix
need to be pulled into cache.  
Yet, there are several avenues open for speeding up our method so that it reaches
state-of-the-art performance. }

\rwchng{Importantly, recent research in computer science has shown that Wilson's algorithm is not
asymptotically optimal for UST sampling. There are recent methods that are
nearly linear in the number of edges, at the cost of a small approximation
error, for instance, \cite{anari2020log,schild2018almost}. However, at this
stage, the asymptotically-optimal algorithms are very complex and, in some
cases, close to unimplementable. Wilson's algorithm, on the other hand, \mrw{is very easy to implement.}}

\rwchng{In terms of practical, short term ideas for speeding up our methods, there are
quite a few options. One is to exploit a multilevel representation. A
second is to use the coupled random forests algorithm of Avena \&
Gaudilli\`ere~\cite{avena_two_2017}, which generates forests for different values of $q$ in one go and
so can produce a regularisation path for the Tikhonov estimator. A third is to
use random forests as preconditioners to CG. We plan to explore these
possibilities in future work.  }

\section{Conclusion}
\label{sec:conclusion}

The Monte Carlo estimators can be used as building blocks in various
problems, and are amenable to theoretical analysis. 
As we have shown, the proposed methods are adaptable to various graph-based optimization algorithms including the generalized SSL framework, label propagation, Newton's method and IRLS. 
\rwchng{Moreover, via a trick known at least since Gremban\mrw{~\cite[ch. 7]{gremban_combinatorial_nodate}} , these RSF techniques can be easily extended to a larger class of matrices than Laplacians, namely symmetric diagonally dominant matrices, as explained for instance in~\cite{kelner2013simple}.
In future work, we will
continue to explore the links between RSFs and graph-related algebra to develop efficient estimators of  graph characteristics such as effective resistances or $\tr(\ma{L}^\dag)$ where $\ma{L}^\dag$ is the Moore-Penrose inverse of $\ma{L}$. }

\appendices
\section*{Acknowledgments}
\mrw{We thank the anonymous reviewers for their comments and suggestions to improve this manuscript.}

\ifCLASSOPTIONcaptionsoff
  \newpage
\fi

\bibliographystyle{IEEEtran}
\bibliography{rsf_journal,IEEEabrv}

\end{document}
